\documentclass{aa}
\usepackage{natbib}
\usepackage{subcaption}
\usepackage{graphicx}
\usepackage{amsmath}
\usepackage{txfonts}
\usepackage{comment}
\usepackage{setspace}
\usepackage{color}

\newcommand{\changed}[1]{#1}      

\begin{document}

\title{The 3D MHD code GOEMHD3 for large-Reynolds-number
astrophysical plasmas}
\subtitle{Code description, verification and computational performance}

\author{J.~Sk\'ala\inst{1,2,4}
        \and
        F.~Baruffa\inst{3}
        \and
        J.~B\"uchner\inst{1}
        \and
        M.~Rampp\inst{3}
}

\institute{Max Planck Institute for Solar System Research, G\"ottingen, Germany,
            \email{skala@mps.mpg.de}
           \and
           Astronomical Institute of Czech Academy of Sciences, Ond\v rejov,
           Czech Republic
           \and
           Rechenzentrum (RZG) der Max Planck Gesellschaft, Garching, Germany
           \and
           University J. E. Purkinje, \'Ust\'i nad Labem, Czech Republic
}


\abstract
   {
   The numerical simulation of turbulence and flows in almost ideal,
   large-Reynolds-number astrophysical plasmas motivates the implementation of
   MHD computer codes with low resistivity. They should be
   computationally efficient and scale well with large numbers of CPU
   cores, allow to obtain a high grid resolution over large simulation
   domains, and be easily and modularly extensible, e.g. to new
   initial and boundary conditions.
   }
   {
    Implementation, optimization and verification of a computationally
    efficient, highly scalable, and easily extensible,
    low-dissipative MHD simulation code for the numerical
    investigation of the dynamics of large-Reynolds-number
    astrophysical plasmas in three dimensions (3D).
   }
   {
    The new GOEMHD3 code discretizes the ideal part of the MHD equations
    using a fast and \changed{ efficient Leap-Frog scheme}  
    which is second-order accurate in space and time and whose initial and
    boundary conditions can easily be modified.
    GOEMHD3 is parallelized based on the hybrid MPI-OpenMP programming
    paradigm, adopting a standard, two-dimensional
    domain-decomposition approach.
   }
   {
   The ideal part of the equation solver is verified by performing
   numerical tests of the evolution of the well understood Kelvin-Helmholtz
   instability and of Orszag-Tang vortices.
   Further it is shown that the computational performance of the code
   scales very efficiently with the
   number of processors up to tens of thousands of CPU cores.
   This excellent scalability of the code was obtained by simulating the 3D
   evolution of the solar corona above an active region (NOAA AR1249) for
   which GOEMHD3 revealed the energy distribution in the solar atmosphere in
   response to the energy influx from the chromosphere through the transition
   region, taking into account the weak Joule current dissipation and
   viscosity in the almost dissipationless solar corona.
   }
   {
   The new massively parallel simulation code GOEMHD3 enables
   efficient and fast simulations of almost ideal,
   large-Reynolds-number astrophysical plasma
   flows, well resolved and on huge grids covering large domains.
   Its abilities are verified by comprehensive set of tests of ideal and weakly
   dissipative plasma phenomena.
   The high resolution ($2048^3$ grid points) simulation of a large part
   of the solar corona above an observed
   active region proves the excellent parallel
   scalability of the code up to more than 30.000 processor cores.
   }


\authorrunning {J. Skala et al.}  
\titlerunning {GOEMHD3 for large-Reynolds-number plasmas}
\maketitle
%

\section{Introduction}
\label{intro}

For most astrophysical plasmas the viscosity and current dissipation 
(resistivity) are negligibly small, i.e. astrophysical plasmas are nearly 
ideal, 
almost dissipationless and hence, for relevant processes and scales, the 
characteristic Reynolds and Lundquist numbers are very large. This requires 
specific approaches to correctly take into account turbulence and different 
kinds of ideal and non-ideal interactions in the plasma flows like, e.g., shock 
waves, dynamo action and magnetic reconnection \citep{BirnPriest2007}. 
Fortunately, improvements in computer technology as well as the development of 
efficient algorithms allow increasingly realistic numerical simulations of the 
underlying space plasma processes \citep{Buchner2003LNP}. For the proper 
numerical description of nearly dissipationless astrophysical plasmas, e.g., of 
magnetic reconnection \citep{Buchner2007} and dynamo action one needs to 
utilize 
schemes with negligible numerical diffusion for MHD as well as kinetic plasma 
descriptions \citep{Elkina2006}. The schemes should be as simple as possible in 
order to run quickly and efficiently. Moreover, in order to ensure flexibility 
concerning the particular physics problem under consideration they should allow 
an easy modification of initial and boundary conditions as well as the simple 
addition and adjustment of physics modules. For this sake, e.g. the serial 
second-order-accurate MHD simulation code LINMOD3D had been developed. It was 
successfully applied to study the magnetic coupling between the solar 
photosphere and corona based on multi-wavelength observations 
\citep{Buchner2004a}, to investigate the heating of the transition region of 
the 
solar atmosphere \citep{BuchnerEtal2004}, and the acceleration of the fast 
solar 
wind by magnetic reconnection \citep{BuchnerNikutowski2005}. It was also used 
to 
physically consistently describe the evolution of the solar chromospheric and 
coronal magnetic fields \citep{BuchnerNikutowski2005a} and for comparing solar 
reconnection with spacecraft telescope observations~\citep{Buchner:2007SolB}, 
the electric currents around EUV bright points~\citep{SantosEtal:2008-1}, the 
role of magnetic null points in the solar corona \citep{Santos++2011} and the 
triggering of flare eruptions \citep{Santos+:2011}. Other typical applications 
of LINMOD3D were the investigation of the relative importance of compressional 
heating and current dissipation for the formation of coronal X-ray bright 
points~\citep{Javadi2011} and of the role of the helicity evolution for the 
dynamics of active regions~\citep{Yang2013}. For the investigation of stronger 
magnetic field gradients in larger regions of the solar atmosphere, however, an 
enhanced spatial resolution is required. To a certain degree this was possible 
using the OpenMP parallelized code MPSCORONA3D which can be run on large shared 
memory parallel computing resources, e.g.\ for the investigation of the 
influence of the resistivity model on the solar coronal heating 
\citep{Adamson+:2013}.

For the simulation of further challenging problems, like the development and 
feedback of turbulence, for high resolution simulations of large spatial 
domains, for the investigation of turbulent astrophysical plasmas with very 
large Reynolds numbers, for the consideration of subgrid-scale turbulence for 
large scale plasma phenomena, one needs to be able to utilize, however, a much 
larger number of CPU cores than shared memory systems can provide. Hence, 
MPI-parallelized MHD codes like, e.g. 
ATHENA~\footnote{https://trac.princeton.edu/Athena}, BATS-R-US~\footnote{ 
http://ccmc.gsfc.nasa.gov/models/modelinfo.php?model=BATS-R-US}, BIFROST, 
ENZO~\footnote{http://enzo-project.org} or 
PENCIL~\footnote{http://pencil-code.nordita.org} have to be used which run on 
distributed memory computers. PENCIL is a sixth-order spatial and third-order 
in 
time accurate code. It uses centered spatial derivatives and a Runge-Kutta time 
integration scheme. ENZO is a hybrid (MHD + N-body) code with adaptive mesh 
refinement which uses a third-order piecewise parabolic method 
\citep{Colella+Woodward:1984} with a two-shock approximate Riemann solver. 
ATHENA allows a static mesh refinement, implementing a higher order scheme and 
utilizing a Godunov method on several different grid geometries (Cartesian, 
cylindrical). It employs third-order cell reconstructions and a Roe solver, 
Riemann solvers as well as a split corner-transport upwind 
scheme~\citep{Colella:1987,Stone+:2008} with a constrained-transport 
method~\citep{Evans+Hawley:1988,Stone+Gardiner:2009}. BIFROST is a code which 
is 
sixth-order accurate in space and third-order accurate in time 
\citep{Gudiksen+:2011}. BATS-R-US solves the 3D MHD equations in finite-volume 
form using numerical methods related to Roe's approximate Riemann Solver. It 
uses an adaptive grid composed of rectangular blocks arranged in varying 
degrees 
of spatial refinement levels. Note that all these codes are of an accuracy 
higher than second order. As a result every time step is numerically expensive 
and changes or modifications, e.g. of initial and boundary conditions require 
quite some effort.
Contrary, second-order-accurate schemes are based on simpler numerics and 
efficient solvers. They are generally far easier to implement, modify, e.g. 
concerning different types of initial and boundary conditions, are parallelize. 
On modern computer architectures the desired numerical accuracy can rather 
easily and computationally cheaply be achieved by enhancing the grid 
resolution. 
This served as the motivation for our new GOEMHD3 code to be based on a simple 
second-order-accurate scheme which is relatively straightforward to implement 
and to parallelize, and which facilitates modification and extension. GOEMHD3 
runs quickly and efficiently on different distributed-memory computers from 
standard PC clusters to high-performance-computing (HPC) systems like the 
"Hydra" Cluster of the Max-Planck-Society at the Computing Center (RZG) in 
Garching, Germany. In order to demonstrate the reach and limits of the code, 
GOEMHD3 was tested on standard problems as well as by simulating the response 
of 
the strongly height-stratified solar atmosphere based on photospheric 
observations using a large number of CPU cores. In section \ref{Basic} the 
basic 
equations solved by the code are described (\ref{equations}), together with 
their discretization and numerical implementation (\ref{Numerical 
implementation}). In section \ref{Parallelization} the hybrid MPI-OpenMP 
parallelization of GOEMHD3 is described. The performance of the code was tested 
with respect to different ideal and non-ideal plasma processes 
(Sect.~\ref{tests}). All tests are carried out using the same three-dimensional 
code. For quasi-2D simulations the number of grid points in the invariant 
direction is reduced to four, the minimum value required by the discretization 
scheme. Section \ref{Kelvin-Helmholtz} presents a test of the hydrodynamic part 
of the code by simulating the well-posed problem of a Kelvin-Helmholtz velocity 
shear instability using the methodology developed by \citet{McNally+:2012} as 
it 
was applied also to test the higher-order codes like PENCIL, ATHENA and ENZO. 
In 
section \ref{Orszag-Tang} ideal MHD limit is tested by simulating vortices 
according to \citet{Orszag+Tang:1979}. In the past, Kelvin-Helmholtz 
instability 
and Orszag-Tang vortex tests have been used also to verify 
total-variation-diminishing schemes \citep{Ryu+:1995}. The possibility of 
numerical oscillations due to the finite difference discretization was 
investigated as in \citet{Wu:2007}. In order to verify the explicit 
consideration of dissipative processes by GOEMHD3 a current decay test was 
performed suppressing others terms in the equations (Sect.~\ref{Current_decay}).
\changed{
The effective numerical dissipation rate of the new code is assessed by a set of 
one-dimensional, Harris-like current sheet \citep[e.g.,][]{Kliem+:2000} 
simulations (Sect.~\ref{Harris_sheet}) and by a fully three-dimensional 
application with solar-corona physics (Sect.~\ref{sec:numDiss}).
}
\changed{
Section~\ref{corona} presents an application of GOEMHD3 to the
evolution of the solar corona in response to changing conditions at the lower 
boundary according to the photospheric plasma and magnetic 
field evolution, and documents the computational performance of the 
code.  
}
The paper is summarized and conclusions for the use of 
GOEMHD3 are drawn in section~\ref{Conclusions}.

\section{Basic equations and numerical implementation}
\label{Basic}

\subsection{Resistive MHD equations}
\label{equations}

For a compressible, isotropic plasma the resistive MHD equations in
dimensionless form read

\begin{eqnarray}
\label{eq:density}
\frac{\partial\rho}{\partial t}+\nabla\cdot(\rho\boldsymbol u) = \chi \nabla^2 
\rho\\
\label{eq:momentum}
\frac{\partial(\rho \boldsymbol u)}{\partial t}+
\nabla\cdot\left[ \rho\boldsymbol u\boldsymbol u+\frac{1}{2}(p+B^2)\boldsymbol I
-\boldsymbol B\boldsymbol B\right] = \nonumber \\ 
-\nu\rho(\boldsymbol{u}-\boldsymbol{u}_0)
+\chi\nabla^2 \rho\boldsymbol{u} \\
\frac{\partial \boldsymbol B}{\partial t} =
\nabla\times(\boldsymbol u\times \boldsymbol B)-
(\nabla \eta)\times\boldsymbol j+\eta\nabla^2 \boldsymbol B
\label{eq:induction}\\
\frac{\partial h}{\partial t}+\nabla\cdot h\boldsymbol{u} =
\frac{(\gamma-1)}{\gamma h^{\gamma-1}}\eta\boldsymbol{j}^2
+\chi \nabla^2 h
\label{eq:pressure}
\end{eqnarray}

where the symbols $\rho$, $\boldsymbol u$, $h$, and $\boldsymbol B$ denote the 
primary variables, density, velocity and specific entropy of the plasma, and 
the 
magnetic field, respectively. The symbol $\boldsymbol{I}$ is the $3\times 3$ 
identity matrix. The resistivity of the plasma is given by $\eta$ and the 
collision coefficient $\nu$ accounts for the coupling of the plasma to a 
neutral 
gas moved around with a velocity $\boldsymbol u_0$. The system of equations is 
closed by an equation of state. The entropy $h$ is expressed via the scalar 
pressure $p$ as $p=2h^\gamma$. Using the entropy as a variable instead of the 
internal energy (here adiabatic conditions are assumed, i.e. a ratio of the 
specific heats $\gamma = 5/3$) then Eq.~(\ref{eq:pressure}) shows that in 
contrast to the internal energy the entropy is conserved in the absence of 
Joule 
and viscous heating. Ampere's law $\boldsymbol j = \nabla\times \boldsymbol B$ 
allows to eliminate the current density $j$. The terms proportional to $\chi$ 
in 
equations (\ref{eq:density}), (\ref{eq:momentum}), and (\ref{eq:pressure}) are 
added by technical reasons as explained in the next section(~\ref{Numerical 
implementation}).

The variables are rendered dimensionless by choosing typical values for a 
length 
scale $L_0$, a normalizing density $\rho_0$ and a magnetic field strength 
$B_0$. 
For the normalization of the remaining variables and parameters the following 
definitions are used: $p_0=\frac{B_0^2}{2\mu_0}$ for a typical (magnetic) 
pressure , $u_{\mathrm A0}=\frac{B_0}{\sqrt{\mu_0\rho_0}}$ for a typical 
(Alfv\'en) velocity, and $\tau_{\mathrm 0}$ for the Alfv\'en crossing time over 
a distance $L_0$, i.e. $\tau_{\mathrm A0} =\frac{L_0}{u_{\mathrm A0}}$. The 
current density is normalized by $j_0=\frac{B_0}{\mu_0L_0}$, the resistivity by 
$\eta_0=\mu_0 L_0 u_{\mathrm A0}$ and the energy by $E_0=B_0^2L_0^2/\mu_0$. For 
simulations of the solar atmosphere typical numerical values are $L_0 = 
5000\,\mathrm{km}$, $\rho_0 = 2\times10^{15}\,\mathrm{m^{-3}}$ and 
$B_0=10^{-3}\,\mathrm{T}$, which yields $p_0=0.7958\,\mathrm{Pa}$, $u_{\mathrm 
A0}=487.7\,\mathrm{km\;s^{-1}}$, $\tau_{\mathrm A0}=10.25\,\mathrm{s}$, $j_0= 
1.59\times10^{-4}\,\mathrm{A\;m^{-2}}$, 
$\eta_0=3.06\times10^6\,\mathrm{\Omega\;.m}$ and 
$W_0=1.99\times10^{13}\,\mathrm{J}$ for the normalizing energy.

\subsection{Numerical implementation}
\label{Numerical implementation}

The resistive MHD equations (Eqs.~\ref{eq:density}--\ref{eq:pressure}) are 
discretized on a three-dimensional Cartesian grid employing a combination of a 
time-explicit Leap-Frog, a Lax, and a DuFort-Frankel finite difference schemes 
\citep[see][]{Press+:2007}. For the conservative, homogeneous part of the MHD 
equations second-order accurate Leap-Frog discretization scheme
\begin{equation}
      \frac{\psi_i^{n+1}-\psi_i^{n-1}}{2\Delta 
t}=-\frac{\psi^n_{i+1}-\psi^n_{i-1}}{\Delta x}
      \label{eq:leapfrog}
\end{equation}
is adopted.

A first-order Lax method is used to start the integration from initial
conditions, i.e. to compute $\psi^n$ from the given initial values
$\psi^{n-1}$, or upon a change of the time step $\Delta t$ (see below).

\changed{
The advantage of the Leap-Frog scheme lies in its low numerical dissipation -- 
in the derivation of the scheme all even derivative terms cancel in the 
expansion and a von Neumann stability analysis shows that there is no amplitude 
dissipation for the linearized system of MHD equations. The full, non-linear 
system in principle exhibits finite dissipation rates, corresponding to 
additional non-linear terms in the von Neumann stability analysis. As shall be 
shown in sections~\ref{Harris_sheet} and \ref{sec:numDiss} the effective 
numerical dissipation rates found with GOEMHD3 are sufficiently small to enable 
simulations of almost ideal, dissipationless, magneto-fluids with very high 
Reynolds number ($Re\sim 10^{10}$).
}
The disadvantage of the Leap-Frog scheme is that it is prone to generate
oscillations. When such numerical oscillations arise they must be damped,
e.g. by a locally switched-on diffusivity which prevents a steepening of
gradients beyond those resolved by the grid. This also prevents mesh drift
instabilities of staggered Leap-Frog schemes which are due to the fact
that odd and even mesh points are decoupled
(see, e.g., \citealt{Press+:2007} and \citealt{Yee:1966}).

This general oscillation-damping diffusion is explicitely introduced via
terms proportional to $\chi\nabla^2 \rho$, $\chi\nabla^2 \rho\boldsymbol{u}$
and $\chi\nabla^2 h$ in the right hand sides of equations
(\ref{eq:density}), (\ref{eq:momentum}), and (\ref{eq:pressure}).
Finite $\chi$ are switched on only when necessary for damping as explained
below.
Hence, although a Leap-Frog scheme is by construction dissipationless  
\changed{ (at least in the linear regime)} we combined it with an explicit, 
externally 
controllable diffusion necessary to avoid oscillations which in the end makes 
the scheme dissipative but in a controlled way. In order to maintain 
second-order accuracy the dissipative terms are discretized by a DuFort-Frankel 
scheme which is also used to discretize the diffusion term in the induction
equation (\ref{eq:induction}):

\begin{equation}
\label{eq:duforFrankel}
\psi_i^{n+1}=\psi_i^{n-1}+2\Delta t \left[ w_1 \psi_{i-1}^n+w_3 
\psi_{i+1}^n+\frac{1}{2}w_2\left( \psi_i^{n-1}+\psi_i^{n+1}\right)\right]
\end{equation}

Here, $w_1=\frac{2}{\Delta x_l\Delta x}$,
$w_2=\frac{-2}{\Delta x_l\Delta x_r}$ and $w_3=\frac{2}{\Delta x_r\Delta x}$
are the coefficients necessary to calculate the second order derivatives on
the non-equidistant mesh, used.
The left derivative is denoted by $\Delta x_l=x_i-x_{i-1}$,  right $\Delta
x_l=x_{i+1}-x_i$ and total $\Delta x=\Delta x_l+\Delta x_r$.

Combining the Leap-Frog (Eq.~\ref{eq:leapfrog}) and the DuFort-Frankel
(Eq.~\ref{eq:duforFrankel}) discretization schemes one obtains
\begin{equation}
  \psi^{n+1}=\psi^{n-1}+\lambda\left[ S^n + \sum_i \left( \chi_i H_i - dx_i 
F_i^n\right) \right]
\label{eq:all}
\end{equation}
with the fluxes $F_i^n$ and source terms $S_i$
\begin{equation}
      \boldsymbol{F}=\left(
  \begin{array}{c}
    \boldsymbol{\rho\boldsymbol{u}}\\
    \rho\boldsymbol{u}\boldsymbol{u}-\boldsymbol{B}\boldsymbol{B}
    +\frac{1}{2}\boldsymbol{I}(p+B^2)\\
    \hat{\boldsymbol \epsilon}_{3\times 3} \cdot \boldsymbol{E}\\
    h\boldsymbol{u}
  \end{array}
  \right) \ ,\quad
  \boldsymbol{S}=\left(
  \begin{array}{c}
     0\\
     -\nu\rho(\boldsymbol{u}-\boldsymbol{u}_0)\\
     -(\nabla \eta)\times\boldsymbol j\\
     \frac{(\gamma-1)}{\gamma h^{\gamma-1}}\eta\boldsymbol{j}^2
  \end{array}
  \right)\,.
  \end{equation}

The diffusion term is $H_i=w_1\psi_{i-1}^n+w_3\psi_{i-1}^n+w_2\psi_i^{n-1}$,
where $\hat{\boldsymbol \epsilon}_{3\times 3}$ is the permutation pseudo-tensor,
$\boldsymbol{E} =-\boldsymbol{u}\times\boldsymbol{B}$ is the convection
electric field, and $\psi$ represents any one of the plasma variables $\rho$,
$\rho\boldsymbol u$, $\boldsymbol B$ and $h$. Eq.~(\ref{eq:all}) further
uses the abbreviations $\lambda=\frac{dt}{1-\frac{1}{2}\sum \chi_i w_{2,i}}$,
$dt=2\Delta t$, $dx_i=\frac{1}{\Delta x_i}$ and the index $i$ represents
the $x$, $y$, and $z$ directions.

Note that two terms of the source vector $\boldsymbol S$ are not
treated exactly according to this scheme:
due to the staggered nature of the Leap-Frog scheme the values of
$h$ at time level $n$ are not available in the pressure equation.
Similarly, for the induction equation, the gradient of the resistivity is
needed at time level $n$.
While in the former case, $h^n$ can simply be approximated by averaging
over the neighboring grid points, now the gradient $\nabla\eta^{n}$ is
extrapolated from the previous time level $n-1$, assuming that the arising
numerical error is small for a resistivity that is reasonably smoothed
both in space and in time. The resistivity is smoothed in time in case a
time dependent resistivity model is used.
 GOEMHD3 is meant to describe collisionless astrophysical plasmas,
e.g. of the solar corona, where resistivity is physically caused by
micro-turbulence.
Since it is not possible to describe kinetic processes like micro-turbulence
in the framework of a MHD fluid-model different kinds of switch-on resistivity
models are implemented in GOEMHD3 to mimic kinetic scale current dissipation
at the macro-scales.
The criteria controlling the switch-on of resistivity usually localize
the resistivity increase. This allows, e.g., to reach the observed magnetic
reconnection rates. Anyway, current dissipation is expected to be most prominent
in regions of enhanced current densities where the use of smooth resistivity
models is appropriate.

As noted before, the numerical oscillations are damped by
switching on diffusion.  As soon as in any of the three coordinate
directions the value of $\psi$ exhibits two or more local extrema
(either maxima or minima) the diffusion coefficient $\chi $ is given
a finite value, here, e.g., $\simeq 10^{-2}$ at the given
grid-point and its next neighbours.
If at least two extrema are found then the diffusion term is switched on
locally in the corresponding direction.
For this all directions (x, y and z) are considered separately.

For solar applications it is possible to start GOEMHD3 with initially force
free magnetic fields.
Such magnetic fields are obtained by a numerical extrapolation of the
observed photospheric magnetic field.
In order to improve the accuracy in case of strong initial magnetic fields
the current density is evaluated by calculating
$\boldsymbol j = \nabla\times (\boldsymbol B - \boldsymbol{B_{init}})$,
i.e. for a field from which the initial magnetic field $\boldsymbol{B_{init}}$
is subtracted.
This reduces the error arising from the discretization of the magnetic field.
In this case the current density is explicitly used to solve the momentum
equation which obtains the form

\begin{equation}
\frac{\partial\rho \boldsymbol u}{\partial t}+\nabla\cdot\left[ \rho\boldsymbol 
u\boldsymbol u+\frac{1}{2}p\boldsymbol I \right]
- \boldsymbol j \times \boldsymbol B
= -\nu\rho(\boldsymbol{u}-\boldsymbol{u}_0)+\chi
\nabla^2 \rho\boldsymbol{u}
\label{eq:momentum2}
\end{equation}
For this sake GOEMHD3 alternatively can solve the momentum equation
Eq.~(\ref{eq:momentum2}) instead of Eq.~(\ref{eq:momentum}).

\paragraph{Time step control}
The time-explicit discretization entails a time-step limit according to the
Courant-Friedrichs-Lewy (CFL) condition, which basically requires that during a
time step no information is propagated beyond a single cell of the numerical 
grid.
To this end the minimum value of the sound, Alfv\'en and fluid crossing times,
and similarly for the resistive time scale, is determined for every grid cell,
\begin{equation}
 \Delta t=\xi\cdot\min_l(\frac{\Delta x_l}{\max_l(c_s,u_A,u)},\frac{\Delta 
x_l^2}{\max_l(4\eta)})
\label{eq:timestep}
\end{equation}
with the local values of the sound speed,
$c_s=\sqrt{{\gamma p}/{\rho}}$,
the Alfv\'en speed, $u_A=\sqrt{{\boldsymbol{B}^2}/{\rho}}$,
and the macroscopic velocity $u=|\boldsymbol{u}|$ at the grid position $l$. 
Typically, a value of 0.2
is chosen for the constant safety factor
$\xi\in(0,1)$.

In our simulations the time step $\Delta t$ is changed only after
several (usually at least $\sim$10) time steps which avoids interleaving a
necessary Lax integration step too frequently and hence compromising the
overall second-order accuracy of the Leap-Frog scheme.
In order to prevent an unlimited decrease of the time step, limiting values
like, e.g., at least\ $10\%$ and $1\%$ of the initial values of the density and 
the
entropy $h$, respectively, and $u<3u_A$ are enforced. The values at the
corresponding grid points are reset to the corresponding cut-off value and the
values at the surrounding grid points are smoothed by averaging over the
neighboring grid points.

\paragraph{Divergence cleaning}

Due to discretization errors unphysical finite divergences of the
magnetic field may arise. In order to remove such finite values of
$\nabla B$ the following cleaning method is applied which
solves a Poisson equation for the magnetic potential $\phi$:
\begin{eqnarray}
 \Delta \phi = \nabla . \boldsymbol{B}^{'}
 \label{Poisson} \\
 \boldsymbol{B} = \boldsymbol{B}^{'} - \nabla \phi
\end{eqnarray}
where $\boldsymbol{B}^{'}$ is the magnetic field with a finite divergence
and $\boldsymbol{B}$ is the cleaned magnetic field. With central differences 
$d_x=1/(x_{i+1}-x_{i-1})$,
 and alike
for the other coordinate direction which are suppressed here for brevity, the 
Poisson equation
Eq.~(\ref{Poisson}) is discretized as
\begin{equation}
d_x(d_{x-1}+d_{x+1})\overline{\phi}_i^{k+1}=d_x(B_{x+1}-B_{x-1})-d_x(d_{x-1}
\phi_{i-2}^k+d_{x+1}\phi_{i+2}^k)
\label{eq:Poissondiscrete}
\end{equation}
and solved with a simple fix-point iteration where $k$ denotes the iteration 
step.
For faster convergence a standard relaxation method is utilized,
\begin{equation}
 \phi_i^{k+1}=\xi \overline{\phi}_i^{k+1}+(1-\xi)\phi_i^k
\end{equation}
where the relaxation coefficient $\xi$ depends on the iteration $k$ as
\begin{equation}
 \xi=\frac{1}{4} 
\left(\tanh\left(\frac{16k}{k_{max}}-2\right)+1\right)+\frac{1}{2}
\end{equation}

\subsection{Hybrid MPI-OpenMP parallelization}
\label{Parallelization}

The time-explicit discretization scheme described above can be straightforwardly
parallelized using a domain decomposition approach and introducing halo
regions (``ghost zones'') of width 1, corresponding to an effective stencil 
length of 3 in each of the
coordinate directions. Accordingly, only next-neighbor communication and a
single global reduction operation (for computing the time step,
cf.~Eq.\ref{eq:timestep}) are
necessary for exchanging data between the domains.
To be specific, GOEMHD3 employs a two-dimensional domain
decomposition in the $y-z$ plane with width-1 halo exchange,
using the Message Passing Interface (MPI). Within the individual,
"pencil"-shaped domains, a shared-memory parallelization is
implemented using OpenMP.
The hybrid MPI-OpenMP approach firstly integrates
smoothly with the existing structure of the serial code and secondly,
thanks to a very efficient OpenMP parallelization within the domains,
allows utilizing a sufficiently large number of processor cores, given
typical sizes of the numerical grid ranging between $256^3$ and $2048^3$
points. In addition, the hybrid parallelization helps to maximize the size 
(i.e.~volume
in physical space) of the individual MPI domains, and hence to
minimize the surface-to-volume ratio. The latter translates
into a smaller communication-to-computation ratio and hence
relatively smaller communication times, and the former
accounts for larger MPI messages and hence decreases
communication overhead (latency).
Our parallelization assumes the individual MPI domains
to be of equal size (but not necessarily with a
quadratic cross section in the $y-z$ plane). This greatly
facilitates the technical handling of the extrapolations required by
so-called line-symmetric side-boundary conditions \citep{Otto+:2007}
which are often employed in realistic solar corona simulations.
As a side effect, this restriction {\it a priori} avoids
load-imbalances due to an otherwise non-uniform distribution of the
processor workload.

Overall, as shall be demonstrated below (cf.\ Sect.~\ref{scalability}),
GOEMHD3 achieves very good parallel efficiency over a wide range of
processor counts and sizes of the numerical grid, with the hybrid
parallelization outperforming a plain MPI-based strategy at high
core counts.

\section{Test problems}
\label{tests}

In order to assess the stability, the convergence properties and the numerical
accuracy of the new GOEMHD3 code, we simulate the standard test
problems of the Kelvin-Helmholtz instability and the
Orszag-Tang vortex, perform a test \citep{Skala+Barta:2012} of
the resistive MHD properties of the code, estimate the effective 
numerical dissipation in the non-linear regime using a Harris-like
current sheet
and compare our results with numerical and analytical reference solutions.
All tests are two-dimensional problems in the space coordinates.
In order to perform such two-dimensional simulations with our
three-dimensional code the $x$-direction is considered invariant
and the numerical grid in this direction covers the minimum number of
four points.

\subsection{Kelvin-Helmholtz instability}
\label{Kelvin-Helmholtz}

The properties of the hydrodynamic limit of the GOEMHD3 code are verified by
simulating the non-linear evolution of the Kelvin-Helmholtz instability (KHI)
in two dimensions. This is a well-known standard test of numerical schemes
solving the equations of hydrodynamics \citep[see, e.g.,][]{McNally+:2012}.
The KHI instability is caused by a velocity shear. At its non-linear stage
it leads to the formation of large-scale vortices.
The time evolution of the size and growth rate of the vortices can be
followed and compared with reference solutions obtained by other numerical
schemes.
We verify GOEMHD3 closely following \citet{McNally+:2012}. These authors
established a standard methodology for the KHI test, published and sent us the
results of their fiducial reference solutions obtained using the PENCIL code
of simulations for high-resolution grids with up to $4096^2$ grid points.
In order to avoid problems of resolving sharp discontinuities that
arise in some numerical schemes, \citet{McNally+:2012} proposed a test
setup with smooth initial conditions as introduced by
\cite{Robertson+:2010}. For the two spatial coordinates $0<y<1$ and $0<z<1$)
the initial conditions are, therefore:

\begin{equation}
\zeta=\left\{
  \begin{array}{ll}
    \zeta_1-\zeta_m \exp\left( \frac{z-1/4}{L}\right) & \mathrm{if}\,1/4>z\geq0 
\\
    \zeta_2+\zeta_m \exp\left( \frac{1/4-z}{L}\right) & 
\mathrm{if}\,1/2>z\geq1/4 \\
    \zeta_2+\zeta_m \exp\left( \frac{z-3/4}{L}\right) & 
\mathrm{if}\,3/4>z\geq1/2 \\
    \zeta_1-\zeta_m \exp\left( \frac{3/4-z}{L}\right) & \mathrm{if}\,1>z\geq3/4
  \end{array}
  \right.
  \label{eq:kelHel-rho0v0}
\end{equation}

where $\zeta$ denotes either the density $\rho$ or the velocity
$u_y$, and $\zeta_m=(\zeta_1+\zeta_2)/2$.
In order to trigger the instability a small perturbation $u_z=0.01\sin(4\pi y)$
is imposed on the velocity in the $z$-direction while the initial pressure
is assumed to be uniform in space: $p=5$.
Periodic boundary conditions are imposed. According to the stability
requirements of our code we impose diffusion quantified by a coefficient
$\chi = 4\cdot 10^{-5}$ in Eqs.~(\ref{eq:density},~\ref{eq:momentum}) and
~(\ref{eq:pressure}).

\begin{figure}
  \centering
  \includegraphics[width=0.99\hsize,angle=0]{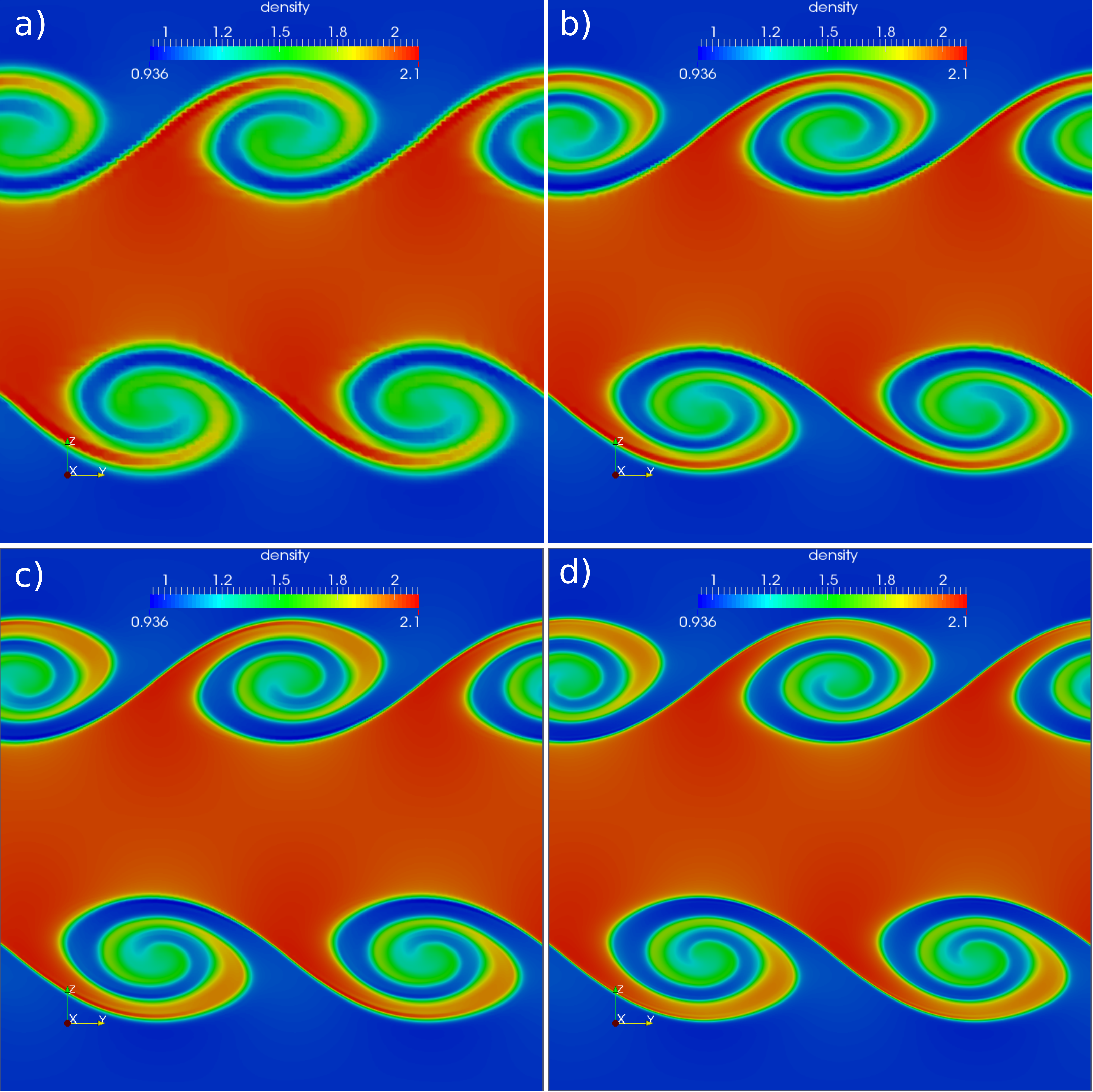}
  \caption{Colour coded mass density, $\rho(y,z)$ at time $t=2.5$ for the
Kelvin-Helmholtz test problem. Panels a), b), c) and d) show the GOEMHD3 results
for a numerical resolution of  $128^2$, $256^2$, $512^2$, and $1024^2$,
respectively.}
  \label{fig:Kelvin2D}
\end{figure}

Figure~\ref{fig:Kelvin2D} shows snapshots of the fluid density at time
$t=2.5$ as computed by GOEMHD3 using different numerical resolution
ranging from $128^2$ (panel a) to $1024^2$ (panel d) grid points.
One recognizes the familiar Kelvin-Helmholtz patterns which qualitatively
compare well with published structures
\citep[e.g.][]{Robertson+:2010,Springel:2010}.
For lower resolutions one observes somewhat smoother edges of the
primary Kelvin-Helmholtz instability which is due to the higher
effective numerical diffusivity caused by the smoothing scheme of
GOEMHD3 (cf.\ Sect.~\ref{Basic}).
For higher resolutions, like $512^2$ (panel c) and $1024^2$ (panel d),
secondary billows develop in the primary billows.
As \citet{McNally+:2012} pointed out, these secondary
billows are artifacts caused by numerical grid noise.

\begin{figure}[t]
  \centering
  \includegraphics[width=0.7\hsize,angle=-90]{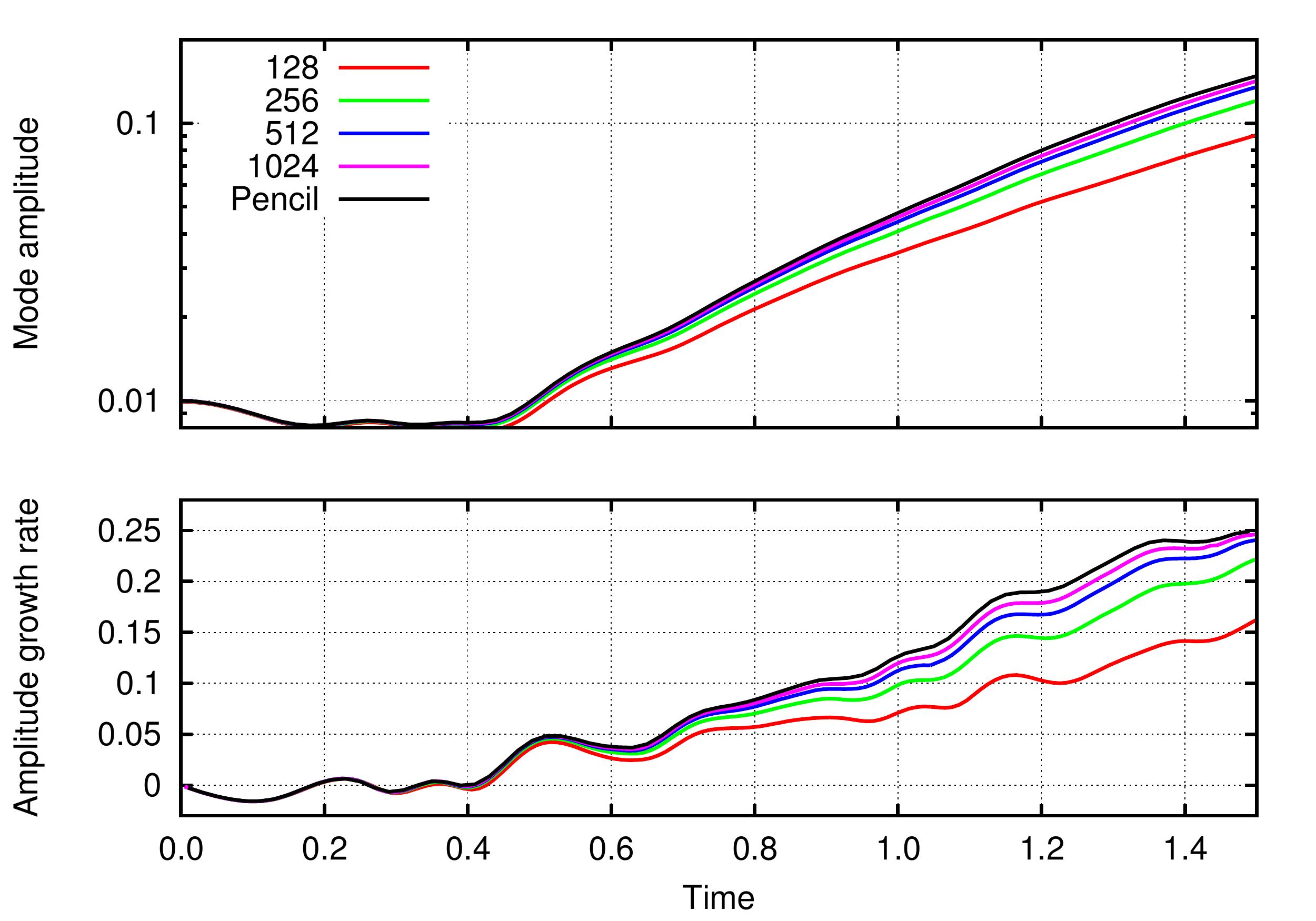}
  \caption{Time evolution of the $y$-velocity mode amplitude
    $A_y$ (top panel) and of its growth rate $\dot A_y$ (bottom panel)
    in the course of the Kelvin-Helmholtz test. The results obtained by GOEMHD3
    for different spatial resolutions are colour coded according
    to the legend.
    The black line corresponds to the result obtained by a PENCIL
    code run using a grid resolution of $4096^2$
    \citep{McNally+:2012}.}
  \label{fig:KelvinMods}
\end{figure}

\begin{figure}
  \centering
  \includegraphics[width=0.7\hsize,angle=0]{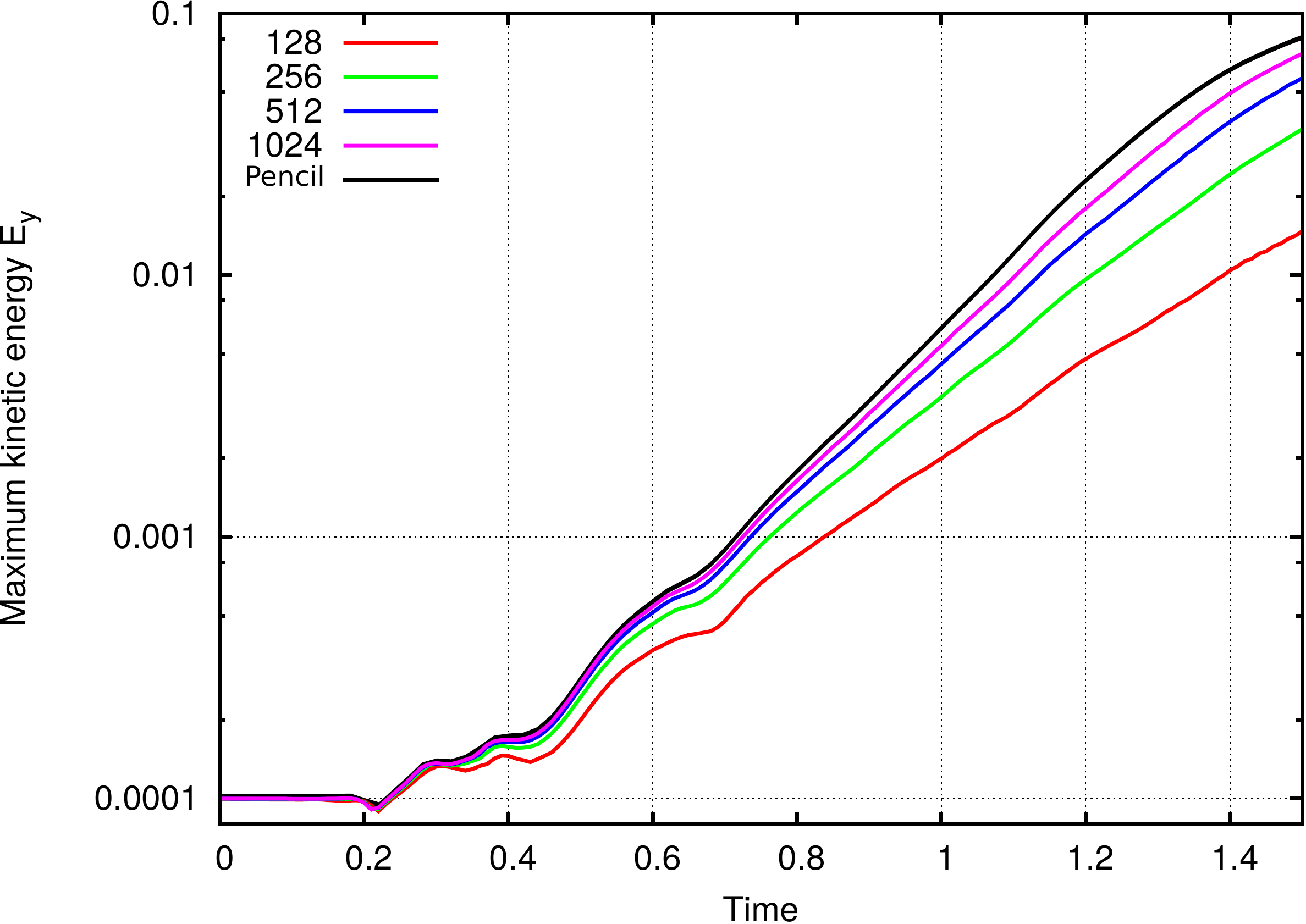}
  \caption{Time evolution of the maximum kinetic energy density
      $E_y=\frac{1}{2}\max_{y,z}(\rho u_y^2)$
    in the Kelvin-Helmholtz test.
    The results obtained by GOEMHD3
    for different spatial resolutions are colour coded according
    to the legend.
    The black line corresponds to the result of a PENCIL code run
    using a grid resolution of $4096^2$ \citep{McNally+:2012}.}
  \label{fig:KelvinKinEn}
\end{figure}

\begin{figure}
  \centering
  \includegraphics[width=0.7\hsize,angle=-90]{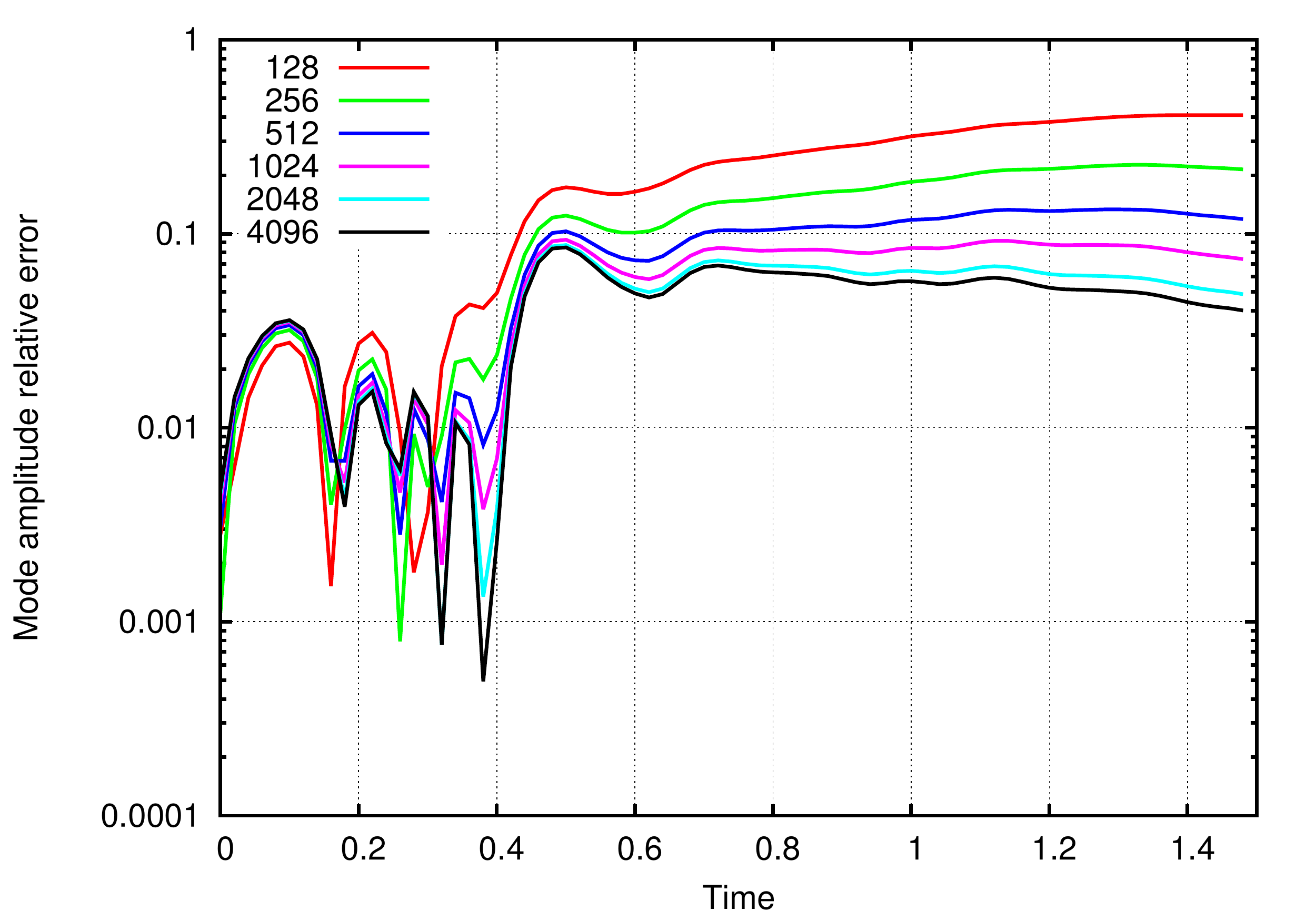}
  \caption{Time evolution of the relative error $|\varepsilon_A|$
  of the mode amplitude obtained for the Kelvin-Helmholtz test
  using GOEMHD3 compared to those obtained by the PENCIL code
  \citep{McNally+:2012} for different numbers of grid points (colour
  coded).}
  \label{fig:errMod}
\end{figure}

\begin{figure}
  \centering
  \includegraphics[width=0.7\hsize,angle=-90]{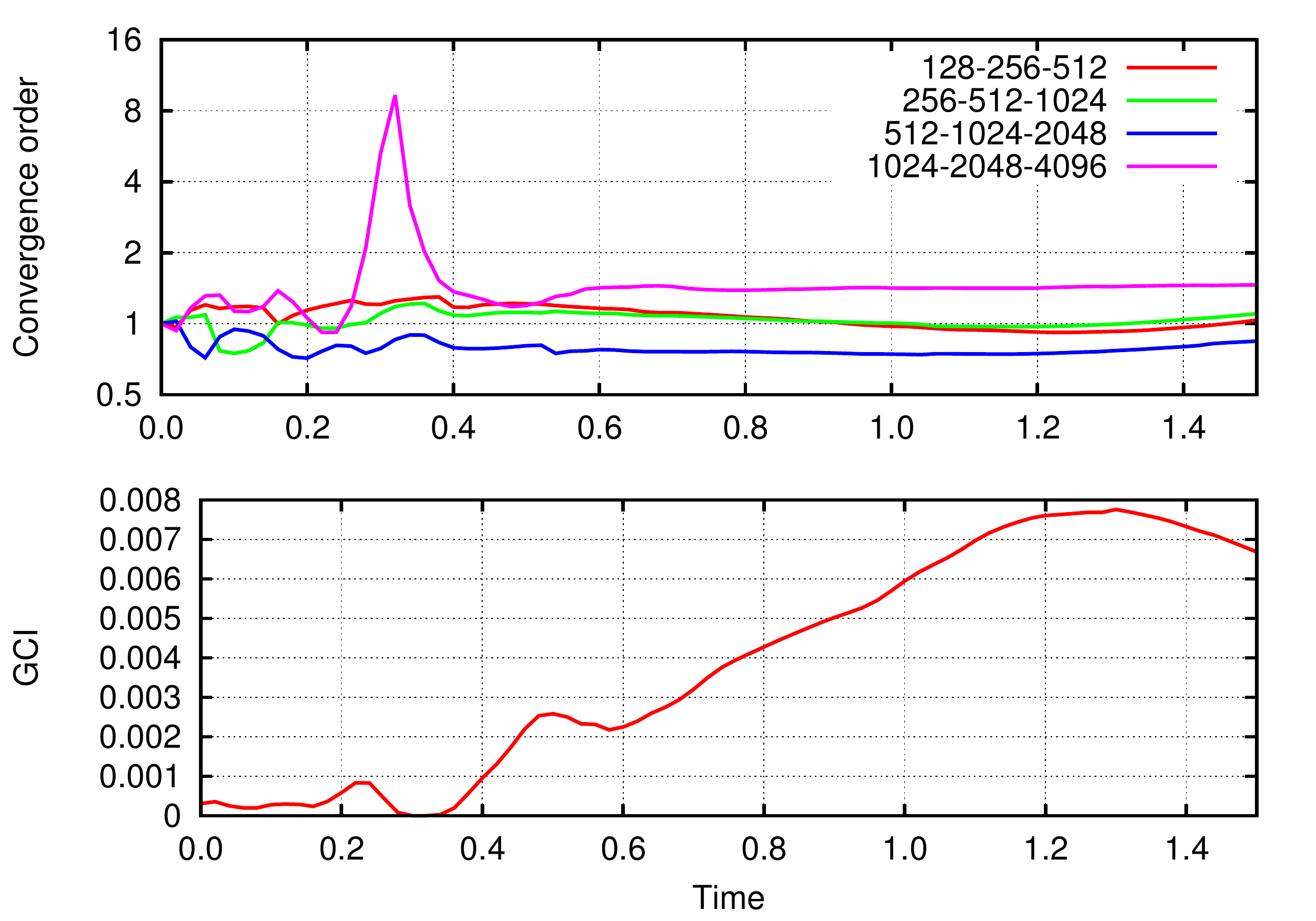}
  \caption{Time evolution of the grid convergence rate (top panel) of
  the mode amplitude in dependence on the spatial resolution given in the
  legend and
  of the grid convergence index GCI (bottom panel) of the mode amplitude
  uncertainty for the highest resolution.}
  \label{fig:GCI}
\end{figure}

For a quantitative verification of GOEMHD3 we compute the time evolution of
different variables introduced and defined by \citet{McNally+:2012}.
First we calculated the $y$-velocity mode amplitude $A_y$ according to
Eqs. (6) to (9) in \citet{McNally+:2012}, its growth rate $\dot A_y$
and the spatial maximum of the kinetic energy density of the motion
in the $y$-direction ($E_y=\frac{1}{2}\max_{y,z}(\rho u_y^2)$).
We further calculated the relative error comparing GOEMHD3 results
with those of the PENCIL-reference code as obtained
by \citet{McNally+:2012} who used the PENCIL code with a grid resolution
of $4096^2$ points. Finally, we calculated convergence quantities as
defined by \cite{Roache:1998} for GOEMHD3.
The results of these calculations are shown in Figures.~\ref{fig:KelvinMods},
\ref{fig:KelvinKinEn}, \ref{fig:errMod}, \ref{fig:GCI}.

First, Figure~\ref{fig:KelvinMods} shows that both, the $y$-velocity mode
amplitude $A_y$ of the KHI (upper panel) and its growth rate
(lower panel) converge well with increasing numerical resolution of
GOEMHD3.
They also exhibit a very good overall agreement with the reference solution
obtained by using the PENCIL code.
A closer look reveals, however that while the initial evolution of $A_y$
closely resembles the reference solution at high as well as at lower
resolution, at later times a sufficiently high resolution of at least
$512^2$ is needed to match the PENCIL code results.
The velocity mode growth rate $\dot A_y$ and the maximum of the kinetic
energy density of the motion in the $y$-direction $E_y$ behave
similarly as one can see in Figure~\ref{fig:KelvinKinEn}. While initially
GOEMHD3 follows the reference solution at all tested resolutions, at later
times GOEMHD3 for lower resolution is slightly smaller than the one obtained
by the PENCIL code.

Further we benchmarked GOEMHD3 by comparing it with the KHI test results
obtained by the PENCIL code for the same initial conditions.
We quantified the comparison by calculating the relative error
$|\varepsilon_A |$ of the mode amplitude $A_y^G$ obtained by
GOEMHD3 with the corresponding values $A_y^P$ obtained by a $4096^2$ grid points
run of the PENCIL code for reference:

\begin{equation}
 \varepsilon_A=\frac{A_y^G-A_y^P}{A_y^P}
\end{equation}

For the whole time
evolution of the KHI until $t=1.5$ (the last value available from
\citealp{McNally+:2012}) Figure~\ref{fig:errMod} shows
the relative errors of the GOEMHD3 results compared to the benchmark solution
which was obtained with the PENCIL code using $4096^2$ grid points.
The relative error decreases from 30\% if GOEMHD3 is using
$128^2$ grid points to less than 4\% if GOEMHD3 uses the same
resolution of $4096^2$ grid
points as the PENCIL code. This is a very good result given that
GOEMHD3 uses a numerically much less expensive second order accurate
scheme compared to a sixth order scheme used in the PENCIL code.

Now, we investigate how the mode amplitude converges with increasing
mesh resolution and establish its uncertainty.
The convergence assessment is based on the generalized Richardson
extrapolation method which allows to extract the convergence rate
from simulations performed at three different grid resolutions  with a
constant refinement ratio \citep{Roache:1998}

\begin{equation}
 p=\ln\left( \frac{f_3-f_2}{f_2-f_1}\right)/\ln(r)
\label{eq:p}
\end{equation}

Here, $r=2$ is the mesh refinement ratio, $f_1$, $f_2$ and $f_3$ are
the mode amplitudes for the fine, medium and coarse mesh, respectively.
From the convergence rate we can calculate the Grid Convergence Index
(GCI, \citet{Roache:1998}) which indicates the uncertainty based on
the grid convergence $p$.

\begin{equation}
 GCI=F_s\frac{|\varepsilon|}{r^p-1}
\end{equation}

where $\varepsilon=(f_2-f_1)/f_1$ is a relative error and $F_s=1.25$
is a safety factor. According to \citep{Roache:1998} these values are
used for grid convergence studies in case of comparing three or more
different resolutions.
Figure~\ref{fig:GCI} shows the time evolution of the grid convergence rate
for the mode amplitude (upper panel) and the GCI corresponding to the
finest resolution (lower panel).
The convergence order of the GOEMHD3 runs appeared to be in the range
$(0.8 - 1.5)$.
A convergence order $p$ of the order of up to 1.5 for GOEMHD3,
a second order accurate code, is a very good results compared to
convergence orders of about 2 obtained by higher (e.g. sixth-) order
accurate schemes like PENCIL.
At the same time the mode amplitude uncertainty GCI for the highest
resolution stays always below 0.008.

The differences between the results obtained by GOEMHD3 and PENCIL at
later times originate from the different role of diffusivity in the codes.
While the Leap-Frog scheme implemented in GOEMHD3 is intrinsically not
diffusive it initially also does not switch on diffusion since no strong
gradients are present which would cause numerical oscillations.
Hence the initial (linear) phase of the Kelvin-Helmholtz instability is
well described by GOEMHD3 since it does not need additional smoothing
at this stage. However, secondary billows develop earlier in the GOEMHD3
KHI test simulation than in PENCIL code simulations
(see Figure~\ref{fig:Kelvin2D} and also Figure~12 of
\citealp{McNally+:2012}).
This is due to the explicit diffusion which is switched on by
the GOEMHD3 code when steep gradients have to be smoothed which develop during
the turbulent phase of the KHI.
As a result GOEMHD3 initially, when it is still not diffusive at all,
reveals the same Kelvin-Helmholtz instability growth despite it is
only second order accurate. Later, however, at the non-linear stage of the
KHI, the explicit diffusion used in GOEMHD3 for smoothing increases above
the diffusion level of the sixth order accurate PENCIL code.

\subsection{Orszag-Tang test}
\label{Orszag-Tang}

\begin{figure}[!t]
  \centering
  \includegraphics[width=0.99\hsize,angle=0]{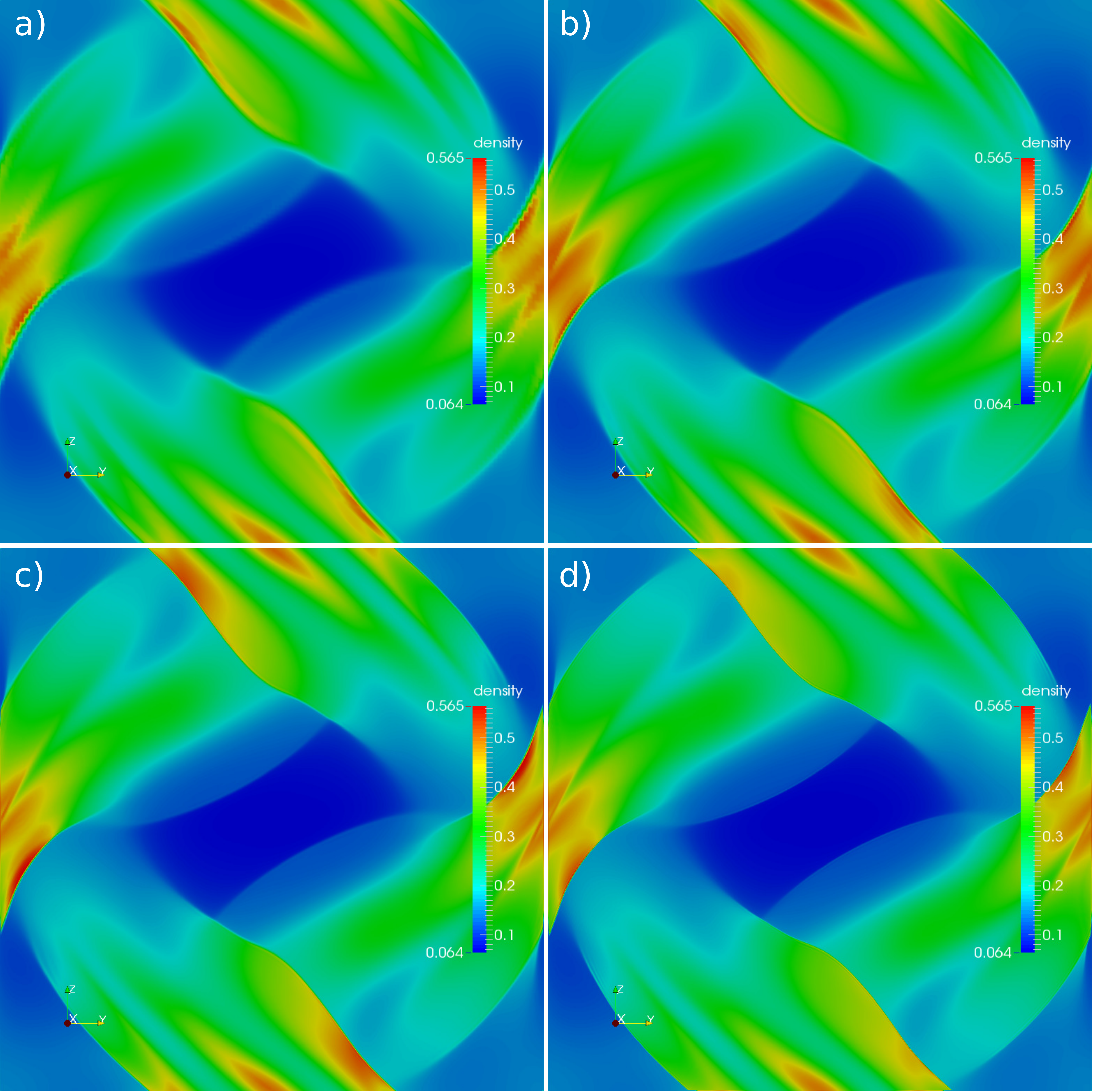}
  \caption{Mass density distribution obtained at $t=0.25$
            for the two-dimensional Orszag-Tang test.
            Panels a), b), c) and d) correspond to a mesh resolution by
            $128^2$, $256^2$, $512^2$ and $1024^2$ grid points,
            respectively.}
  \label{fig:Tang2D-Rho25}
\end{figure}

The ideal-MHD limit of the GOEMHD3 code is tested by simulating a
Orszag-Tang vortex setup in two dimensions \citep{Orszag+Tang:1979}.
The test starts with initially periodic velocity and magnetic
fields, a constant mass density and pressure distribution given by

\begin{eqnarray}
 \rho u_y &=& \sin(2\pi z),\;\; \rho u_z = -\sin(2\pi y)\,, \nonumber \\
 B_y &=& \frac{1}{\sqrt{4\pi}}\sin(2\pi z)\,,\;\;
 B_z = \frac{1}{\sqrt{4\pi}}\sin(4\pi y) \,, \nonumber \\
 \rho &=& \frac{25}{36\pi}\;, \nonumber \\
 p &=& \frac{5}{6\pi}\,. 
\end{eqnarray}

\begin{figure*}[!t]
  \centering
  \includegraphics[width=0.99\hsize,angle=0]{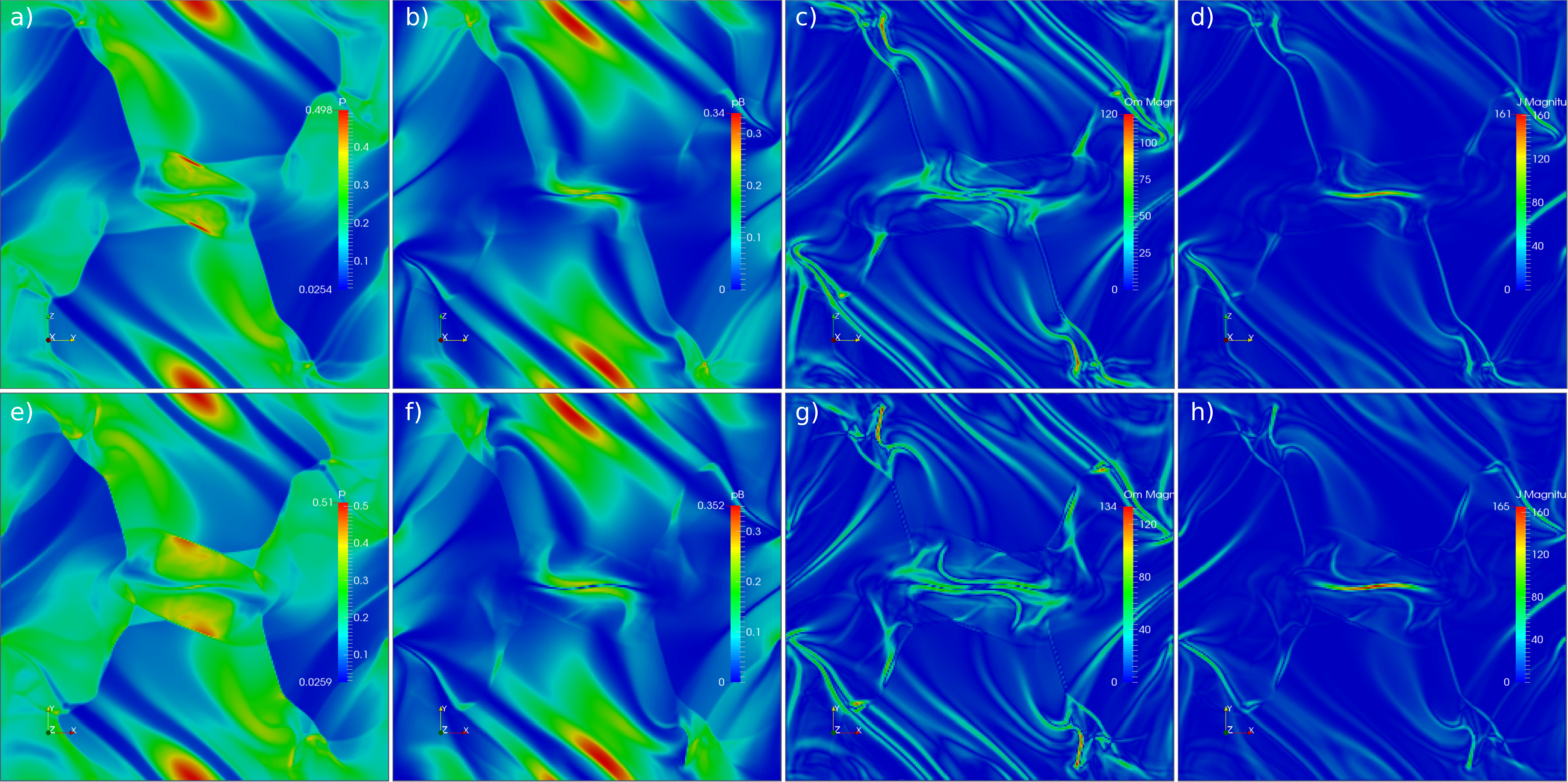}
  \caption{Thermal plasma pressure (panels a, e), magnetic pressure
  (panels b, f), vorticity $\nabla \times \mathrm{u}$ (panels c, g)
  and current density $\nabla \times \mathrm{B}$ (panels d, h)
  obtained for the two-dimensional Orszag-Tang test by GOEMHD3
  (top row) and ATHENA 4.2 (bottom row) for a grid resolution
  of $256^2$. $t=0.5$.}
  \label{fig:Tang2D-PPb}
\end{figure*}

Hence both the velocity and magnetic fields contain X-points, where the
fields vanish. In the $y$-direction the modal structure of the magnetic field
differs from the velocity field structure.
The simulation box size is $\left[-0.5,0.5\right]\times\left[-0.5,0.5\right]$.
All boundary conditions are periodic. The coefficients $\chi$ of the
smoothing diffusion terms are chosen to be $2\times 10^{-4}$,
$1\times 10^{-4}$, $5\times 10^{-5}$ and $2\times 10^{-5}$ for meshes
with $128^2$, $256^2$, $512^2$ and $1024^2$ grid points, respectively.
As expected the GOEMHD3 code reproduces purely growing vortices including
sharp gradients, structures and a dynamics that resembles the results
obtained by \citet{Ryu+:1995} and \citet{Dai+Woodward:1998}.
To give an example Figure~\ref{fig:Tang2D-Rho25}
depicts the mass density distribution at $t=0.25$.
Panels a), b), c) and d) of Figure~\ref{fig:Tang2D-Rho25} correspond to mesh
resolutions of $128^2$, $256^2$, $512^2$ and $1024^2$ grid points, respectively.
Low density regions are colour coded blue, higher density values are red.
Similar structures containing sharp gradients (shocks) were obtained
also, e.g., by ATHENA 4.2 (see, e.g. {\it
http://www.astro.virginia.edu/VITA/ATHENA/ot.html})
and by our least square finite element code
\citep{Skala+Barta:2012}. Note that, owing to its flux conservative
discretization scheme, GOEMHD3 is
able to accurately reproduce the position of shock fronts
(cf.\ Figure~\ref{fig:Tang2D-PPb}).

\begin{figure*}[!t]
  \centering
  \begin{subfigure}{.33\textwidth}
    \includegraphics[bb=0 0 427 365,width=.99\textwidth,angle=0]{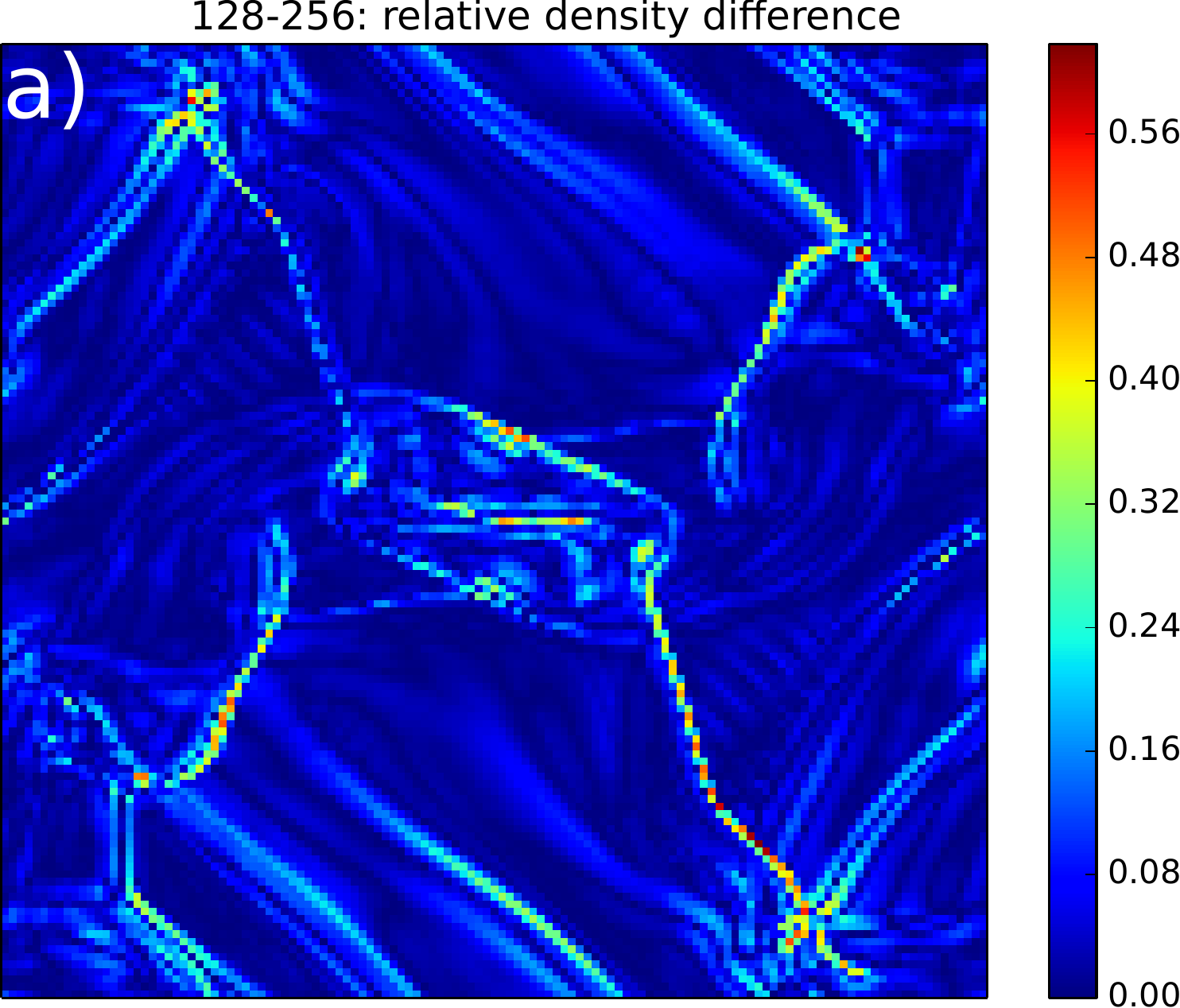}
  \end{subfigure}
  \begin{subfigure}{.33\textwidth}
    \includegraphics[bb=0 0 427 365,width=.99\textwidth,angle=0]{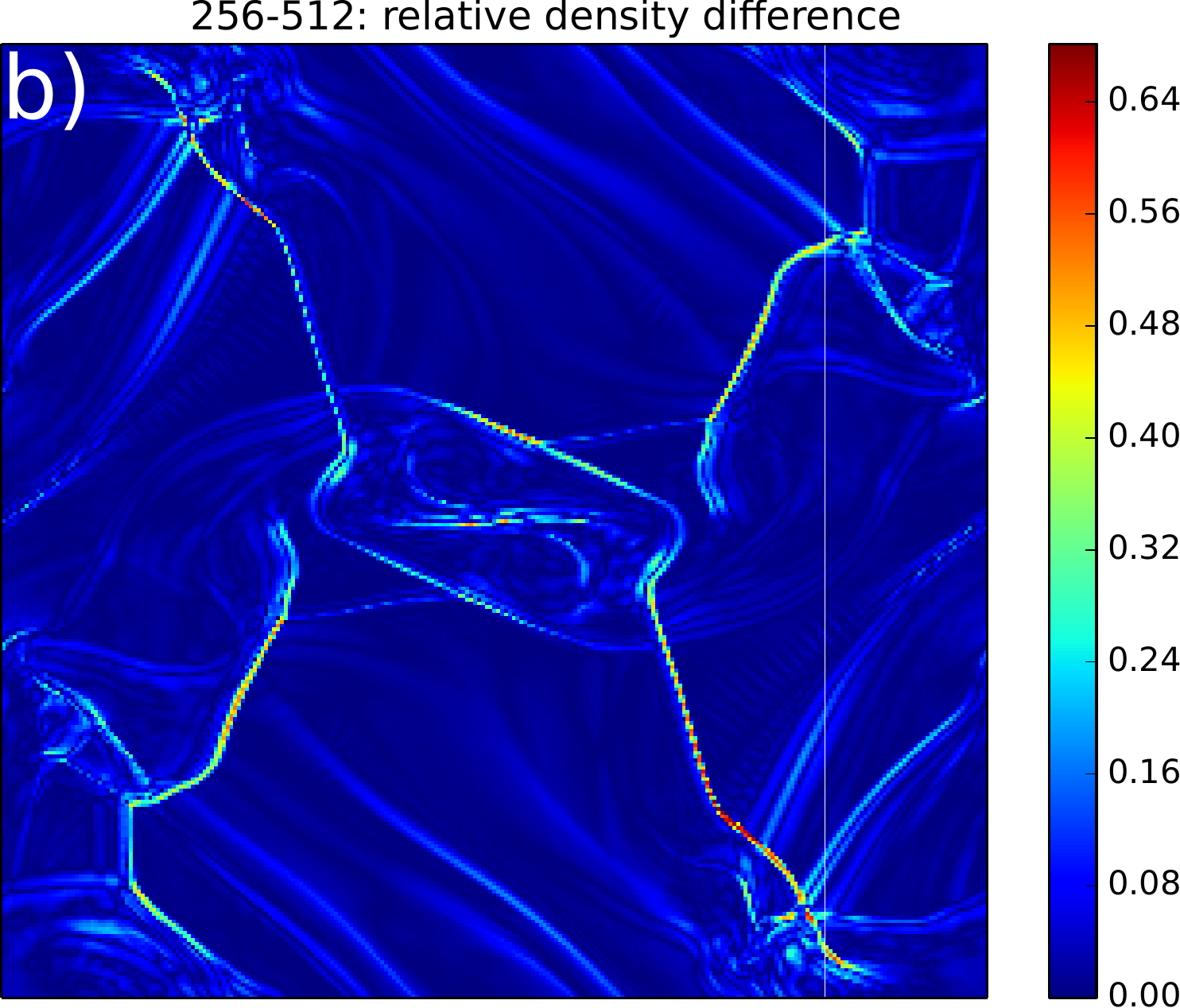}
  \end{subfigure}
  \begin{subfigure}{.33\textwidth}
    \includegraphics[bb=0 0 427 365,width=.99\textwidth,angle=0]{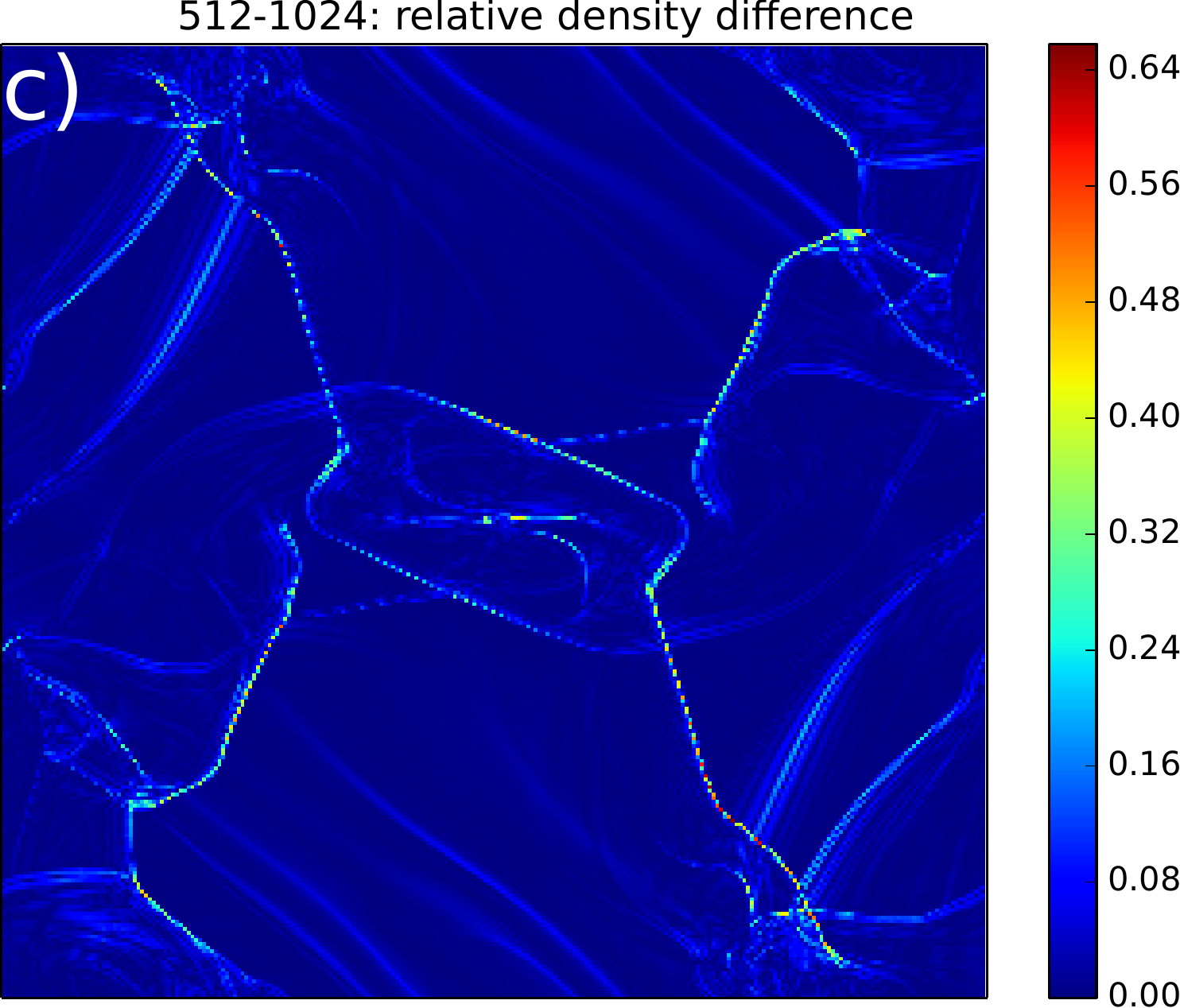}
  \end{subfigure}
  \caption{Spatial distribution of the relative difference $|\varepsilon_\rho|$
    of the mass density obtained by GOEMHD3 simulating the Orszag-Tang vortex
    problem comparing the results for three different mesh resolutions of
    a) $128^2-256^2$, b) $256^2-512^2$ and c) $512^2-1024^2$ grid points.
   }
  \label{fig:Tang2D-DenRelErr}
\end{figure*}

The convergence properties of GOEMHD3 are illustrated by calculating the
relative difference $\varepsilon_\rho=(\rho_2-\rho_1)/\rho_1$ of the spatial
distribution of the mass density obtained by comparing the mass densities
get from runs with lower and higher mesh resolution.
Here $\rho_1$ corresponds to the mass density distribution obtained
for the higher mesh resolution and $\rho_2$ to the coarser grid.
In particular Figure ~\ref{fig:Tang2D-DenRelErr} shows the spatial
distribution of the relative differences obtained at $t=0.5$ for runs with
doubled numbers of grid points - from $128^2$ to $256^2$, from $256^2$ to
$512^2$ and from $512^2$ to $1024^2$ in panels a) to c), respectively.
As one can see in Figure ~\ref{fig:Tang2D-DenRelErr} the largest
relative differences $\varepsilon_\rho$ of the mass density are
localized in regions of strong gradients (shock fronts) while they do not
extend into regions of smooth flows.

In order to directly compare the Orszag-Tang test of GOEMHD3 with the results
of another astrophysical MHD code we have run the same test also using the
ATHENA code in its version 4.2. For this sake we employed the same setup
as described before, a Courant safety constant C = 0.5 and a resolution of
$256^2$ grid points.

Figure~\ref{fig:Tang2D-PPb} compares the Orszag-Tang test simulation results
of GOEMHD3 (top row) with those obtained by running it using the ATHENA 4.2
code(bottom row).
The Figure depicts the two-dimensional spatial distribution of the thermal
pressure (panels a, e), of the magnetic pressure (panels b, f), of the
vorticity $\nabla \times \mathrm{u}$ (panels c, g), and of the current
density proportional to $\nabla \times \mathrm{B}$ (panels d, h)
obtained at $t=0.50$ for a mesh resolution of $256^2$ grid points.
As Figure~\ref{fig:Tang2D-PPb} clearly shows the thermal pressure,
depicted in panels a) an e) and the magnetic pressure (panels b and f) are
anticorrelated everywhere except in the post-shock flows. The comparison
with the ATHENA results shows that the numerically much less expansive
code GOEMHD3 reproduces the ATHENA results leaving just slightly shallower
gradients because of only small diffusion added to smooth gradients
only slightly in order to keep the simulation stable.

As already discussed before GOEMHD3 code switches on a finite diffusion
in order to smooth numerically caused oscillations which may arise due
to the use of a Leap-Frog discretization scheme.
In addition GOEMHD3 limits mass density and pressure to certain
externally given minimum values in order to avoid too large information
propagation (sound and Alfv\'en) speeds which would require very small
time steps to fulfill Courant-Friedrich-Levy condition.
In order to avoid this, the values of the local mass density and the
pressure are replaced by externally prescribed minimum values as soon
as they are reached.
At the same time the values of mass
density and pressure in the neighboring zones of the grid are locally
smoothed towards the minimum value.
Of course, the limiting parameters have to be carefully chosen in a way
to avoid numerically caused local changes of thermal and kinetic energy.

\begin{figure}[!t]
  \centering
  \includegraphics[width=0.7\hsize,angle=-90]{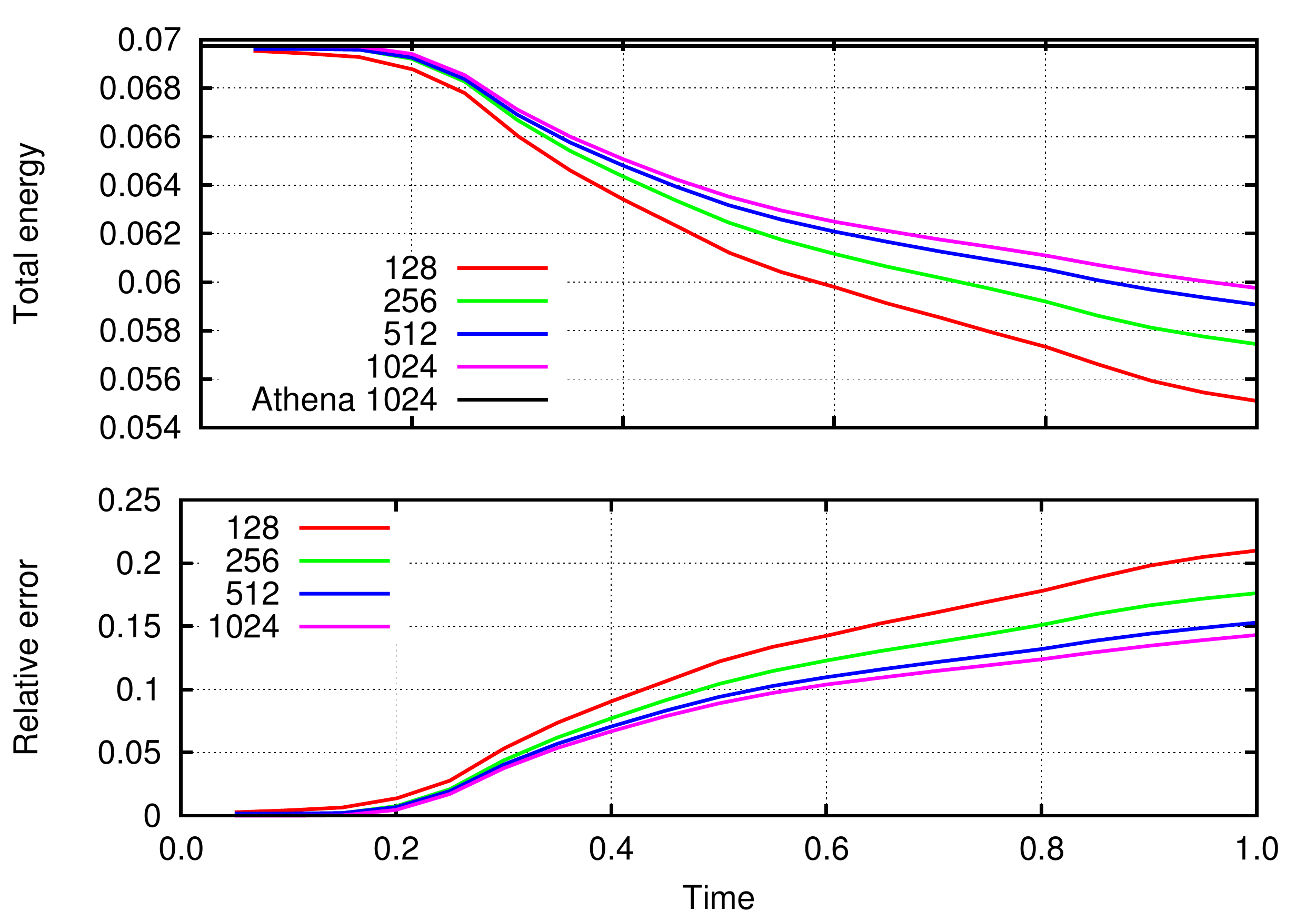}
  \caption{Time evolution of the simulated total energy (upper panel)
  and its relative deviation from its conserved value (lower panel) as
  obtained by the two-dimensional Orszag-Tang test using GOEMHD3 in
  dependence on the mesh resolution of $128^2$ (red), $256^2$ (green),
  $512^2$ (blue) and $1024^2$ (magenta) grid points).
  The black line on top of the upper panel corresponds the rescaled
  from surface to the volume integrated total energy $~0.0697$
  obtained by the ATHENA code run for $1024^2$ grid points.
    }
  \label{fig:Tang-totEng}
\end{figure}

The resulting properties of GOEMHD3 concerning total energy
conservation are documented in Figure~\ref{fig:Tang-totEng} which
shows the resolution-dependent time evolution of the total
energy (upper panel) and of the relative deviation from the conserved
energy (lower panel). GOEMHD3 simulations were performed with resolutions of
$128^2$ (red line), $256^2$ (green line), $512^2$ (blue line) and
$1024^2$ (magenta line) grid points, respectively.
The black line on top of the upper panel corresponds to the
volume-integrated total energy value of $~0.0697$ which is obtained with the
the ATHENA code on a mesh of $1024^2$ grid points (the energy density
on the ATHENA mesh was rescaled from a surface density to a volume
density in order to make it comparable with the three-dimensional GOEMHD3 
simulation).

The coloured curves show the resolution-dependent amount of energy
dissipation of GOEMHD3 - in contrast with (by construction of the numerical
scheme) perfectly energy conserving ATHENA code simulations.
As one would expect, Figure~\ref{fig:Tang-totEng} shows that the energy loss
in GOEMHD3 simulations can be easily reduced by enhancing the numerical
resolution.

\subsection{Resistive decay of a cylindrical current}
\label{Current_decay}

GOEMHD3 was developed to simulate current carrying astrophysical
plasmas taking into account current dissipation.
So, for example, for the description of solar flares magnetic
reconnection has to be simulated, which needs resistivity.
For this sake usually a locally increased resistivity is assumed
which is switched on after reaching, e.g., a macroscopic current
density threshold. After that the resistivity further is linearly
or non-linearly growing with the current density
\citep[see, e.g.,][]{Adamson+:2013}.
For this purpose GOEMHD3 contains modules for spatial and also temporal
smoothing of the resistivity which keeps the simulations stable.
In order to test the ability of GOEMHD3 to correctly describe the
behavior of a resistive magneto-plasma we tested it by applying different
models of resistivity and comparing the simulation results with
analytical predictions where possible.
In particular we applied a test setup simulating the resistive decay
of a cylindrical current column in two spatial dimensions for which in
certain limits analytical solutions exist \citep{Skala+Barta:2012}.

Initially, at $t=0$, a cylindrical current is set up using a
radial magnetic field $\boldsymbol{B}=(0,B_{\phi},0)$, which is given by

\begin{equation}
B_{\phi}(r,t)=j_0\frac{r_0}{x_N} {\mathrm J}_1(x_N \frac{r}{r_0}) \exp(-\alpha 
t)
\end{equation}

in the internal ($r\le r_0$) region and by

\begin{equation}
B_{\phi}(r,t)=j_0\frac{r_0}{x_N} {\mathrm J}_1(x_N)
\end{equation}

in the outer space ($r > r_0$).
Here $j_0=1$ is the amplitude of the current density on the axis of the
cylinder, and $r_0=1$ is the radius of the current column, ${\mathrm J}_l(x)$
denotes a Bessel function of the order $l$, and
$x_N\approx 2.40$ is its first root ${\mathrm J}_0(x)$.
The decrement (current decay rate) $\alpha$ is defined as
$\alpha=\eta(x_N/r_0)^2$.
The pressure is chosen uniformly ($p=1$) in the whole domain and
the density is set to a very large uniform value ($\rho=10^{32}$) which
effectively sets the plasma at rest. Then the system of MHD equations
(\ref{eq:density}--\ref{eq:pressure}) reduces to the induction equation
(\ref{eq:induction}) which in special cases can be solved analytically.
For the GOEMHD3 test simulations the computational domain is chosen
as $\left[-2.5,2.5\right]\times\left[-2.5,2.5\right]$ and open boundary
conditions are applied in the $y$ and $z$ directions.
Periodic boundary conditions are used in the invariant ($x$-)
direction.

We simulated the consequences of resistivity for the evolution
of electrical currents by using GOEMHD3 considering the decay
of a current column in response to two different resistivity models
-- a sharp step-function like and a smooth change of the resistivity.

\paragraph{Step function model of resistivity}

In this model the resistivity was set $\eta=0.1$ in the internal region
while in the outer space it is set to zero.
For such step function of the resistivity distribution \citet{Skala+Barta:2012}
found an analytic solution of the induction equation describing the
time-dependent evolution of the magnetic field and current in the cylinder.
According to this solution the current decays exponentially and
an infinitesimally thin current ring is induced around the resistive
region.
according to

\begin{equation}
\label{eq:jt_anal}
j_x(r,t)=j_0 {\mathrm J}_0(x_N \frac{r}{r_0}) \exp(-\alpha t)
   +\frac{j_0}{2\pi x_N} {\mathrm J}_1(x_N)\left[ 1-\exp(-\alpha 
t)\right]\delta(r-r_o)\ ,
\end{equation}

where $\delta(x)$ is the Dirac delta function.
Due to the discretization of the equations instead of a Dirac delta
function shape the current ring has a finite width which, in our case,
extends over two grid-points while the magnitude of the current
inside this ring is finite.

\begin{figure}
  \centering
  \includegraphics[width=0.7\hsize,angle=-90]{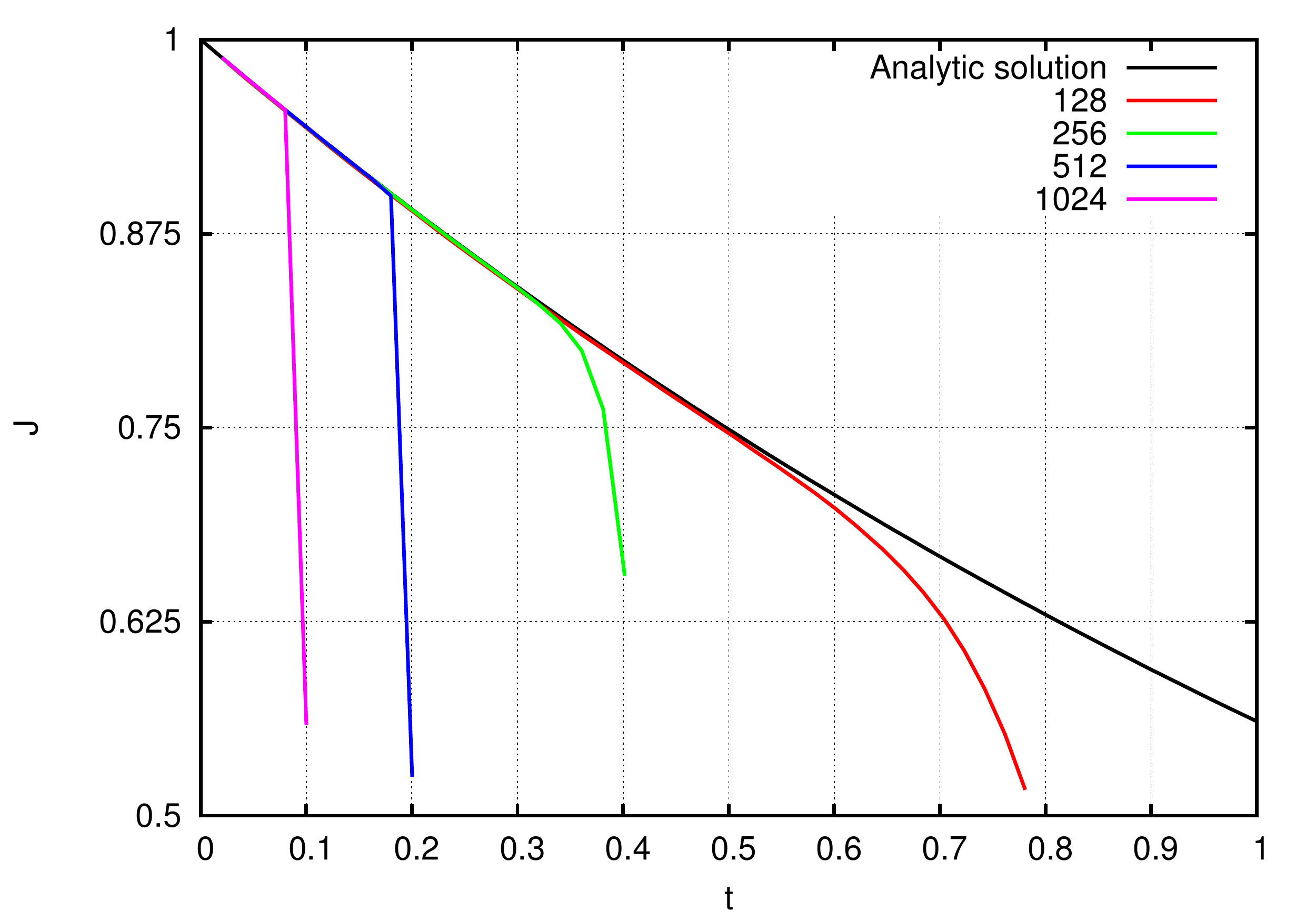}
  \caption{GOEMHD3 simulation of the time evolution of the current density
  $j_x$ at the centre of a current cylinder assuming a step-function change
  of the resistivity. Coloured lines correspond to different mesh resolutions
  employed for the simulations. The solid black line corresponds to the
  analytic solution given by Eq.~(\ref{eq:jt_anal}).}
  \label{fig:curretDecay}
\end{figure}

\begin{figure}[t]
  \centering
  \includegraphics[width=0.99\hsize,angle=0]{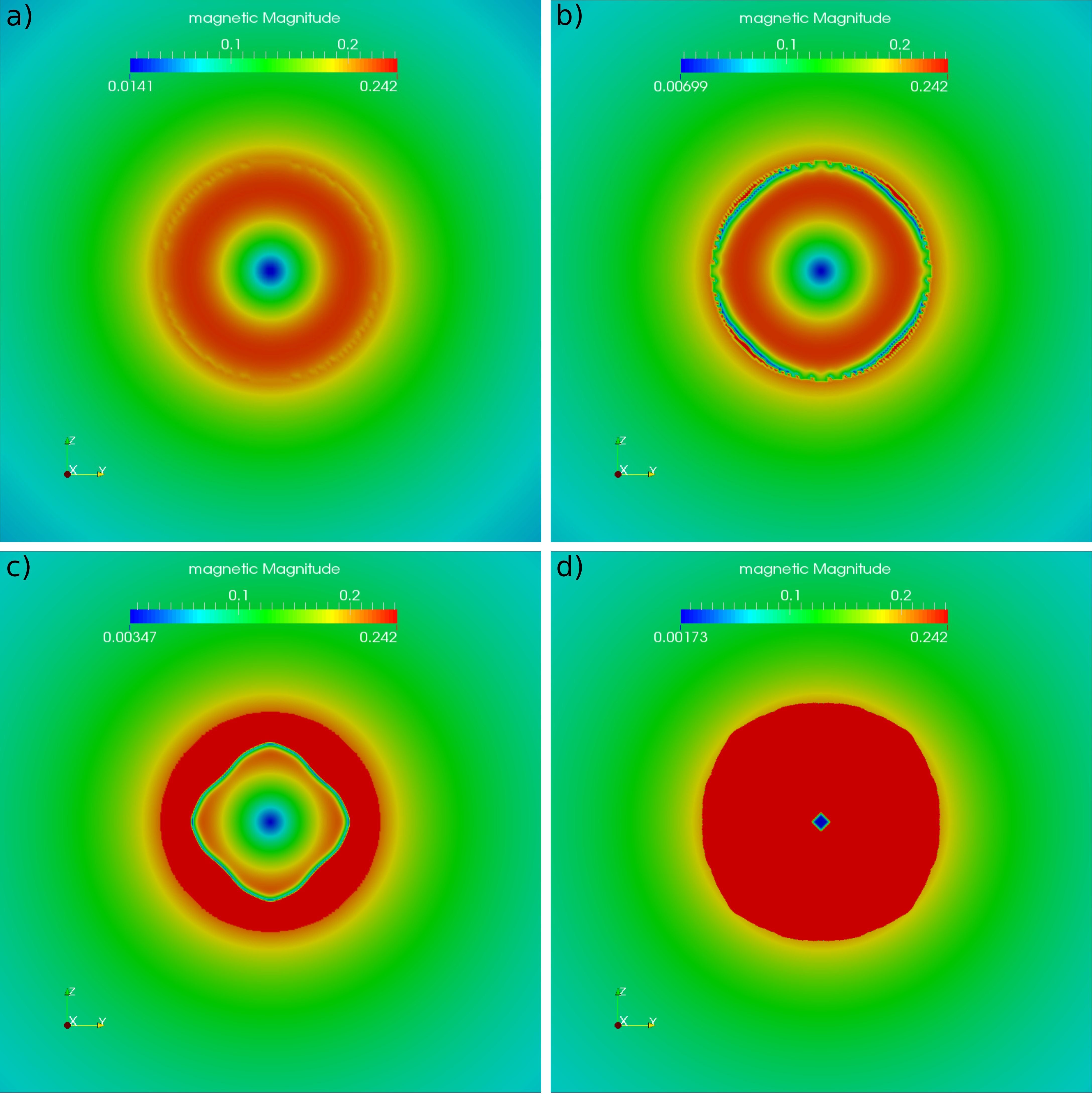}
  \caption{Magnetic field strength $|\boldsymbol{B}|$ in the $y-z$-plane
      at time $t=0.1\tau_A$ for the step-function like change of the
      resistivity changing with mesh resolution.
      Panels a), b), c) and d) correspond to a resolution by
      $128^2$, $256^2$, $512^2$ and $1024^2$ grid points, respectively.
}
  \label{fig:curret2D}
\end{figure}

Figure~\ref{fig:curretDecay} shows that, initially, the decay of the current
density in the center closely follows the time evolution of the analytic
solution (Eq.~\ref{eq:jt_anal}), while a sharp  drop to zero is observed
at later times depending on the numerical resolution of the grid.
As one can see in Figure~\ref{fig:curretDecay} the drop of the current density
at the center of the column is steeper and occurs earlier the better the
grid resolution is. This is due to a numerical instability which spreads
starting from the sharp edge of the resistive cylinder propagating
toward its center.
The growth rate and speed of propagation of this instability increases with
the grid resolution as illustrated by Figure~\ref{fig:curret2D}.
The Figure shows the magnetic field strength $|\boldsymbol{B}|$ in the
$y-z$-plane at time $t=0.1\tau_A$ for four different mesh resolutions
corresponding to $128^2$, $256^2$, $512^2$ and $1024^2$ grid points.
This dependence on resolution is a strong hint at the numerical origin of the
instability to be caused by the sharp resistivity in this model.
To verify this hypothesis we further tested another model in which the
resistivity changes not by a step-function like jump but smoothly, as
it is usually encountered in astrophysical applications.

\paragraph{Smooth change of resistivity model}

\begin{figure}[!t]
  \centering
  \includegraphics[width=0.7\hsize,angle=-90]{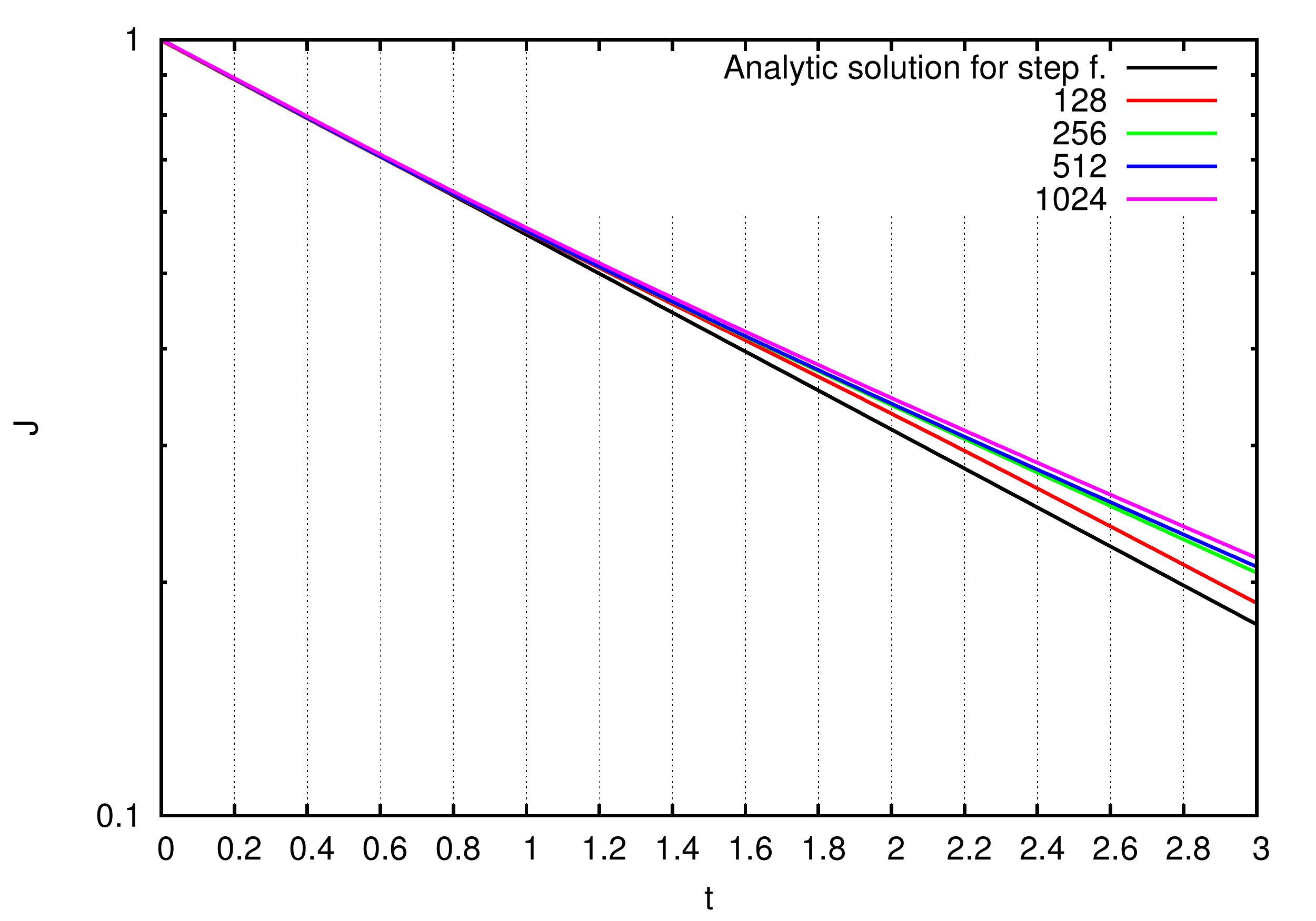}
  \caption{GOEMHD3 simulation of the time evolution of the current density
  $j_x$ at the centre of a current cylinder assuming a smooth change
  of the resistivity. Coloured lines correspond to different mesh resolutions
  employed for the simulations. Note that, since no analytical solution exists
  for this problem the solid black line still corresponds to the analytic
  solution for a step-funtion like change of the resistivity as given by
  Eq.~(\ref{eq:jt_anal}) - as in Figure~\ref{fig:curretDecay}.
  }
  \label{fig:curretDecaySm}
\end{figure}

Indeed, GOEMHD3 is meant to treat collisionless astrophysical plasmas,
like that of the solar corona, by a fluid approach while the resistivity
(as other transport properties) is due to micro-turbulence, not
described by the MHD equations.
As a good compromise usually smoothly changing resistivity models
are assumed to include this microphysics-based phenomenon in the fluid
description.
Smoothly changing switch-on models of resistivity are well suited to mimic
the consequences kinetic scale processes.
To test the influence of a resistivity changing smoothly in space and
time we use the same setup as described in the previous paragraph
just replacing the step-like jump function by a smooth resistivity
change according to

\begin{equation}\label{eq:etasmooth}
 \eta(r)=\eta_0\frac{1}{2}\left( 1-\tanh(\sigma(r-r_0))\right)
\end{equation}

where now $\eta_0=0.1$ and $\sigma$ is a smoothness parameter.
Figure~\ref{fig:curretDecaySm} shows the results obtained for a
smoothness parameter $\sigma = 32$.
It indicates that a smooth resistivity change immediately solves the
problem of oscillatory instabilities arising in case of a step-function
like resistivity change.
As there is no analytical solution known for the smooth switch-on
resistivity we show in Figure~\ref{fig:curretDecaySm} (by a black line)
also the result of the analytical prediction obtained for a step-function
like change of the resistivity.
As one can see in the Figure the simulated current decay is very similar
if compared to the analytically predicted one for the step-function
like jump of the resistivity
The slight deviation of the curves from the predicted one at later times
is perhaps due to the smaller resistivity values arising in the smooth
model at the edge of the resistive cylinder ($r\rightarrow r_0$)
compared to those typical for the step-function model.
Note that the steepness parameter $\sigma = 32$ in Eq.~(\ref{eq:etasmooth})
allowed a stable simulation of the current decay already for the relatively
coarse mesh resolution of $128^2$ grid points as shown in
Figure~\ref{fig:curretDecaySm}. By additional test runs (not shown here) we
tested the stability of the simulations for even steeper resistivity changes and
found that GOEMHD3 can easily cope with changes characterized by
steepness parameters 64, 128 and higher, as long as the grid resolution is 
increased accordingly.

Finally, we conclude that by testing different models of changing resistivity
we could demonstrate that GOEMHD3 can simulate the consequences of localized 
resistive dissipation with sufficient accuracy, provided the changes are
not step-function like.

\changed{
\subsection{Harris current sheet}
\label{Harris_sheet}
In order to assess the effective numerical dissipation rate for the Leap-Frog 
scheme in the non-linear regime a simulation of the Harris-like current sheet 
in 
the framework of an ideal plasma is performed \citep[see, 
e.g.,][]{Kliem+:2000}. 
The size of the simulation box is set to 
$\left[-10.0,10.0\right]\times\left[-0.6,0.6\right]$ with open boundary in 
$y$-direction and periodic boundary conditions in $z$-direction. The initial 
conditions read
\begin{eqnarray}
\rho &=& 1\,,\;\; u_y = u_z = 0\,, \nonumber \\
B_y &=& 0\,, \;\; B_z = \tanh(y)\,, \\
 p &=& 1.01 - \tanh^2(y)\,, \nonumber
\end{eqnarray}
and the physical resistivity is $\eta=0$.

\begin{figure}[!htp]
\centering
\includegraphics[width=0.7\hsize,angle=-90]{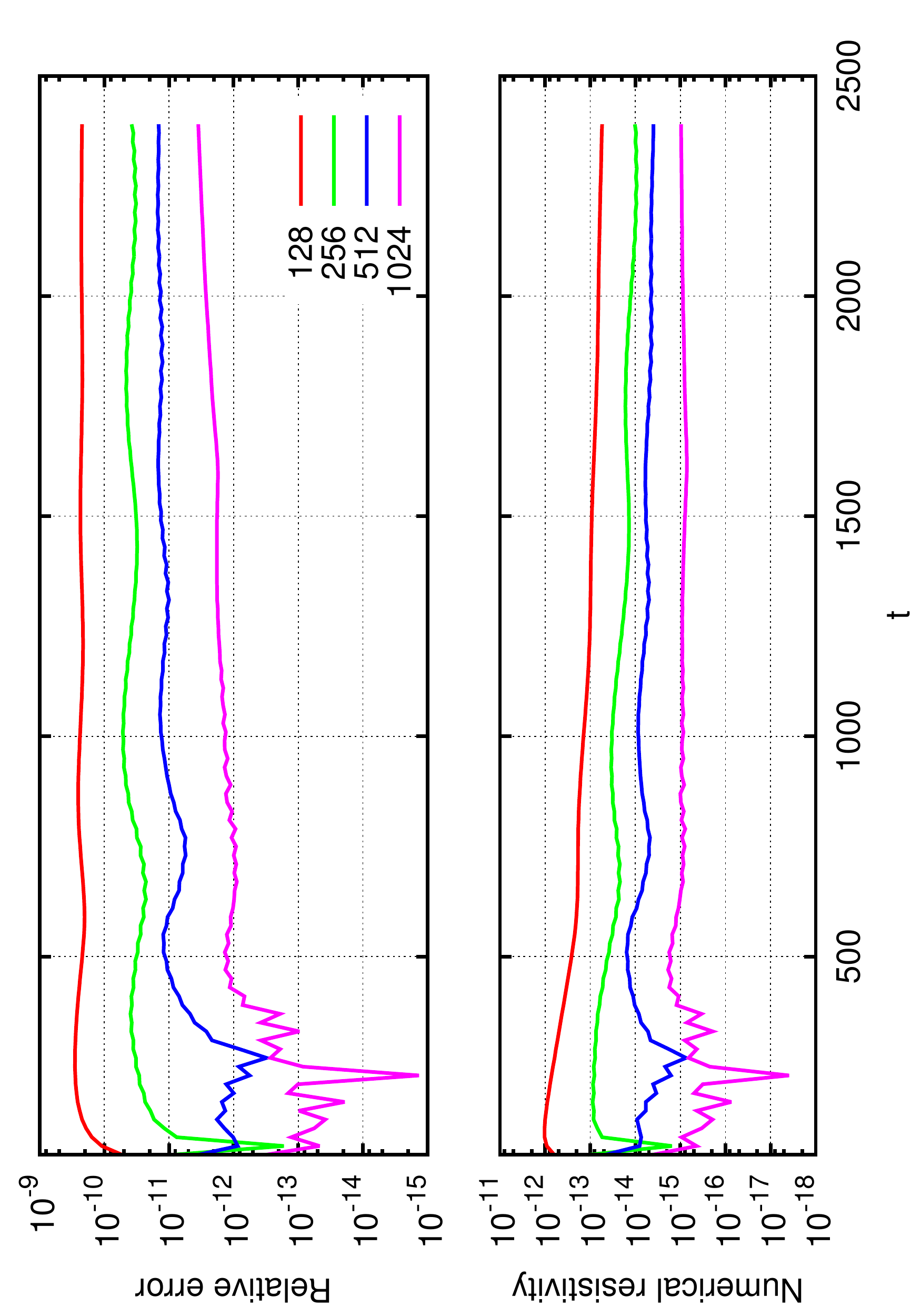}
\caption{Time evolution of the relative deviation $|\Delta B_z/\tanh(y_m)|$ of 
the magnetic field from the analytical prediction (top) and derived effective 
numerical resistivity $\eta_n$ (bottom) at position $(y_m,z_m)=(-0.5493,0)$ for 
the simulation of a Harris-like current sheet. Results for different spatial 
resolutions (number of grid points in $y$ direction) are colour coded according 
to the legend. The high-frequency oscillations are caused by the mesh drift 
instability of the staggered Leap-Frog scheme which is not explicitly damped in 
this simulation setup (cf.\ Sect.~\ref{Numerical implementation}).}
  \label{fig:numDiss}
\end{figure}

We measure the time-variation of the magnetic field, $B_z$, at point 
$(y_m,z_m)=(-0.5493,0)$ where the field attains half of its maximum magnitude. 
The effective numerical resistivity of the discretization scheme can then be 
estimated by
\begin{equation}\label{eq:harris_eta}
 \eta_n=\frac{\Delta B_z}{\Delta t}\cdot\left(\frac{\partial^2 B_z}{\partial 
y^2}\right)^{-1}
\end{equation}
where $\Delta B_z=B_z(y_m,z_m)-\tanh(y_m)$ is the difference between the 
numerical and the analytical solution for the magnetic field, $\Delta t$ 
represents the time of the measurement and the second derivative of $B_z$ is 
approximated by a standard finite-difference representation. 
Figure~\ref{fig:numDiss} shows the time evolution of the numerical dissipation 
rate, $\eta_n$, of the code for different values of the mesh resolution in $y$ 
direction (a constant number of 8 grid points is used in the invariant $z$ 
direction). The relative numerical error of $B_z$ and hence the estimate of the 
numerical resistivity settle at very small values, e.g., $\eta_n\simeq 
10^{-14}$ 
and $|\Delta B_z/\tanh(y_m)| \simeq 5\times 10^{-11}$ for the simulation with 
$256\times8$ grid points. 

We conclude that the residual intrinsic numerical dissipation of the 
discretization scheme is negligible compared with the physical resistivities 
and 
explicit numerical stabilization measures that typically apply in simulations 
with GOEMHD3.
Further below this idealized, one-dimensional test shall be complemented by 
estimates for the effective numerical dissipation rate obtained in fully 
three-dimensional simulations of an eruptive solar region (see 
Sect.~\ref{sec:numDiss}).
}

\section{Three-dimensional simulations of the Solar corona with GOEMHD3}
\label{corona}

In order to demonstrate the applicability of the GOEMHD3 code to
realistic, three-dimensional simulations of weakly collisional astrophysical
plasmas at high Reynolds numbers and to assess the computational performance
of the code we have performed a simulation of the evolution of the
solar corona above an active region.
Being able to simulate such scenarios, where a number of important dynamical 
processes
are still not well understood, has in fact been the main motivation for
developing GOEMHD3. As shall be shown below,
GOEMHD3 allows us to numerically tackle such problems with significantly higher 
numerical
resolution and accuracy as compared with its predecessor codes.

\subsection{Physical context}

We choose for this demonstration the Solar corona above
active region NOAA AR 1429 in March 2012.
This active region is well known since it released many prominent phenomena,
like strong plasma heating, particle acceleration and even eruptions.
Many of them took place during the two weeks between 2$^{nd}$ and 15$^{th}$,
2012 making AR1429 one of the most active regions during the 24$^{th}$
solar cycle.
As a result the morphology of AR 1429 has been thoroughly investigated by a
number of researchers so that the activity phenomena of AR 1429 are now
well known, as they were observed in very details using, e.g., the
AIA instrument on board of NASA's Solar Dynamics Observatory SDO
(see, e.g. \citet{Inoue+:2014}, \citet{Driel-Gesztelyi+:2014},
\citet{Mostl+:2013}).
Very sensitive information was obtained, e.g. about MeV energy (relativistic)
electron acceleration processes which is provided by $30 \, THz$ radio waves.
Examining the role of the continuum below the temperature minimum with a new
imaging instrument operating at El Leoncito \citet{Kaufmann+:2013}
studied the $30 \, THz$ emissions. For the M8 class flare on March 13, 2012,
e.g., they found a very clear $30 \, THz$ signature, much cleaner than
the white-light observations are able to provide.
Another important information about the solar activity are the dynamic
spectra of solar proton emissions. The PAMELA experiment, e.g., measures the
spectra of strongly accelerated protons over a wide energy range.
For four eruptions of AR 1429 the observed energetic protons spectra were
analyzed by \cite{Martucci+:2014}. They interpreted them as an indication
of first order Fermi acceleration, i.e., of a mirroring of the protons
between dynamically evolving plasma clouds in the corona above AR 1429.
Changes in the chemistry of the Earth's atmosphere after the impact of the
energetic solar protons emitted by AR 1429 were studied by
\citet{Clarmann+:2013}. These authors used the MIPAS spectrometer on board
the late European environmental satellite ENVISAT to measure temperature
and trace gas profiles in the Earth atmosphere. They found that the amount
of produced by energetic Solar protons from AR 1429 were among the 12 largest
Solar particle events, i.e. proton storms, in 50 years.
These and more observations of AR 1429 indicate that very efficient energy
conversion processes took place in the corona.

\subsection{Initial and boundary conditions}
\label{conditions}

\begin{figure}[!htp]
\centering
\includegraphics[width=0.5\hsize,angle=-90]{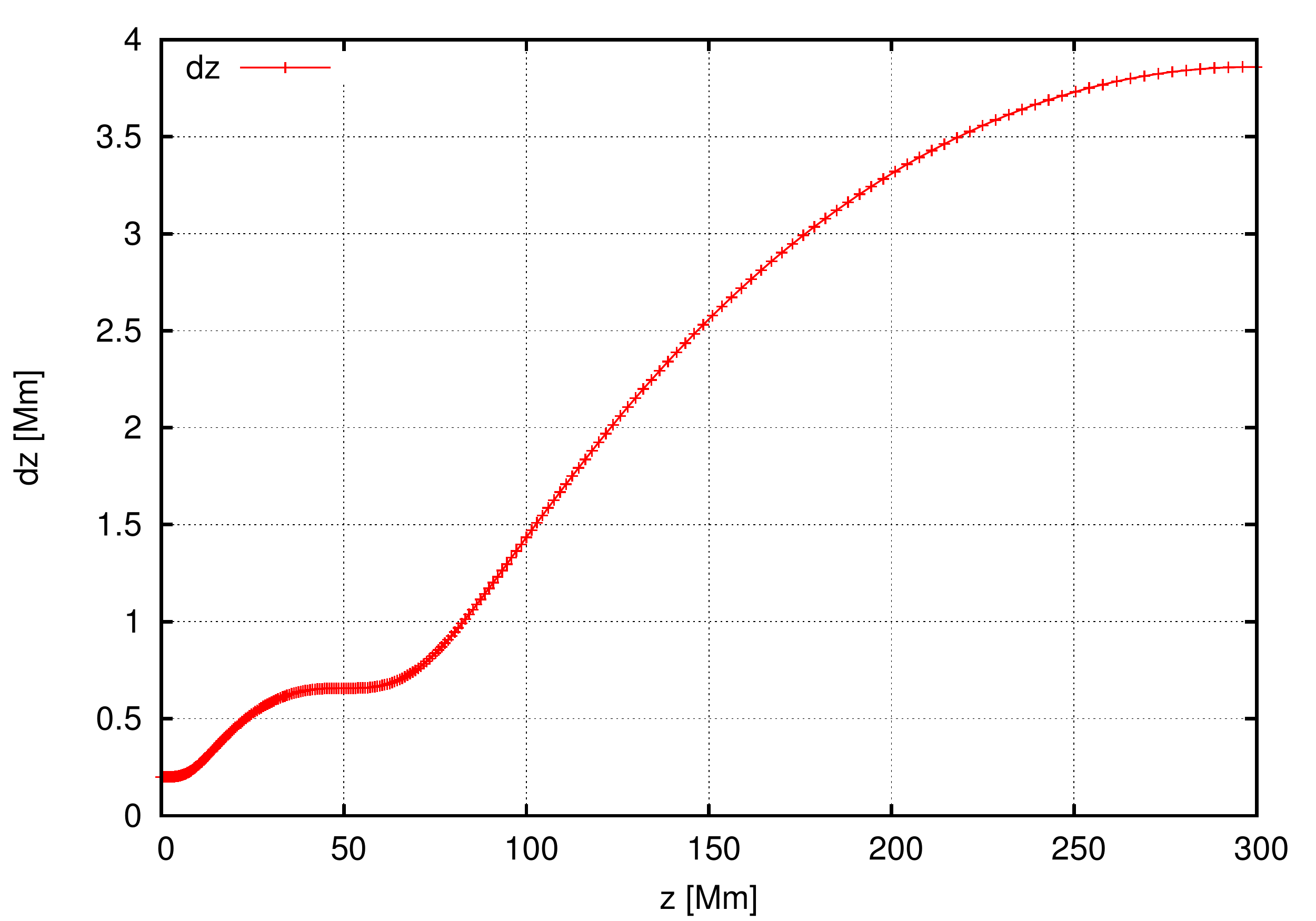}
\caption{The grid spacing $dz$ in the $z$-direction in the simulation of the AR 
1429. Where $z=0$ is the photosphere. The finer spacing at the bottom
part sample better transition region with steep gradients in the density and 
temperature.}
  \label{fig:z-dz}
\end{figure}

We start the simulation with initial conditions derived in accordance with
observations of AR 1429 on March 7$^{th}$ 2012 when at 00:02 UT a X5.4 flare
eruption took place at heliographic coordinates N18E31.
In order to describe the evolution of the corona before the eruption, we
initialize the simulation using photospheric magnetic field observations
on March 6$^{th}$ at 23:35 UT.
Figure \ref{fig:magnetogram} shows the line-of-sight (LOS) component of the
photospheric magnetic field of the AR 1429 obtained at this time by the
HMI instrument on board the SDO spacecraft in a field of view of
300 $\times$ 300~arcsec$^2$. This field of view covers an area of 217.5
$\times $217.5~Mm$^2$ which we choose as the lower boundary of the
simulation box.
The line-of-sight magnetic field is preprocessed by flux balancing,
removing small scale structures and fields close to the boundary before
it is used for extrapolation into 3D. In particular a spatial 2D Fourier
filtering of the  magnetic field data is applied to remove short spatial
wavelength modes with wave numbers greater than 16, which correspond to
structures do not reach out into the corona, above the transition region.
The Fourier filtered magnetic fields are flux balanced
and extrapolated into the third dimension according to the MHD box boundary
conditions derived by \citet{Otto+:2007}. The resulting initial magnetic
field is depicted in Figure~\ref{fig:3Dmagnetic}.
For the height of the simulation box we choose 300~Mm.
The simulation grid spacing in the $x$ and $y$ directions is homogeneous
with a mesh resolution given by the sampling over $258^2$ grid points.
After the filtering out of all modes with wave numbers larger than 16, such
grid allows to resolve all magnetic field structures sufficiently well.
Though in the height ($z$-) direction also $258$ grid points are used
the grid is nonuniformly distributed in order to better resolve the lower
part of the corona / transition region and chromosphere.
Figure~\ref{fig:z-dz} shows the height-dependent grid spacing ($dz$) used.

\begin{figure}[!htp]
\centering
\includegraphics[width=0.5\hsize,angle=-90]{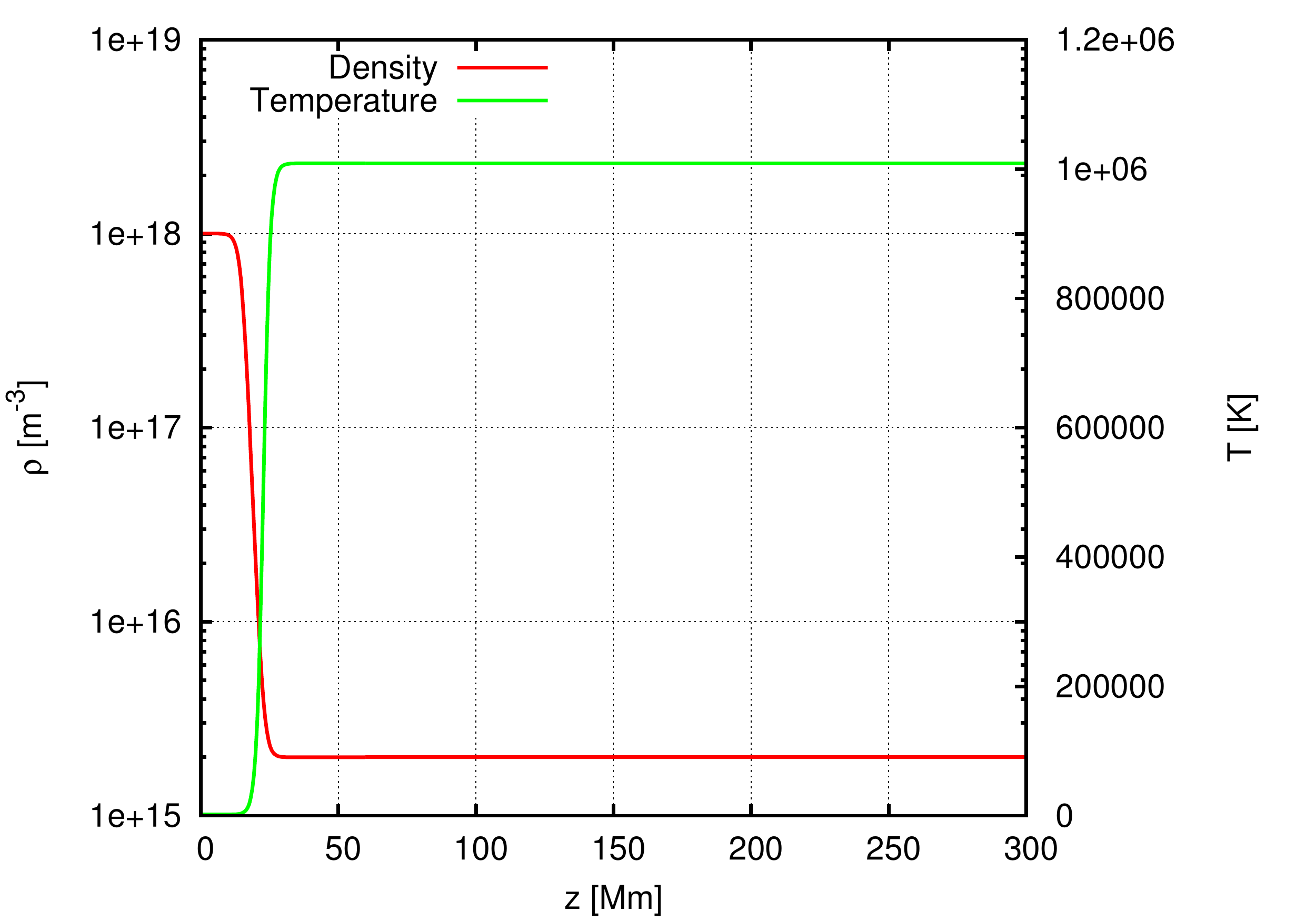}
\caption{Initial height profiles of density and temperature in the simulation 
of 
the AR 1429. The chromospheric density is 500 times larger than the density in
the corona. The transition region is initially localized around $z_0=3$, which 
corresponds to 15 Mm.}
  \label{fig:TemDens}
\end{figure}

\begin{figure}[!htp]
\centering
\includegraphics[width=0.7\hsize,bb=0 0 412 408]{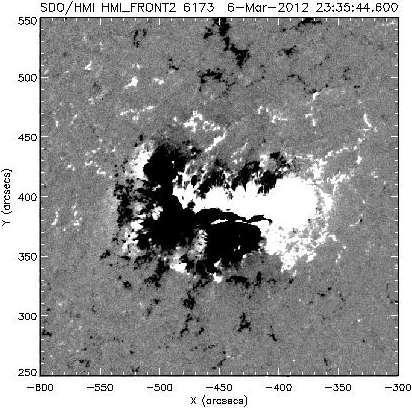}
\caption{Magnetogram of active region 1429 on March 6$^{th}$ 2012
as taken by HMI on board SDO.}
   \label{fig:magnetogram}
\end{figure}

\begin{figure}[!htp]
\centering
\includegraphics[width=0.7\hsize,bb=0 0 803 803]{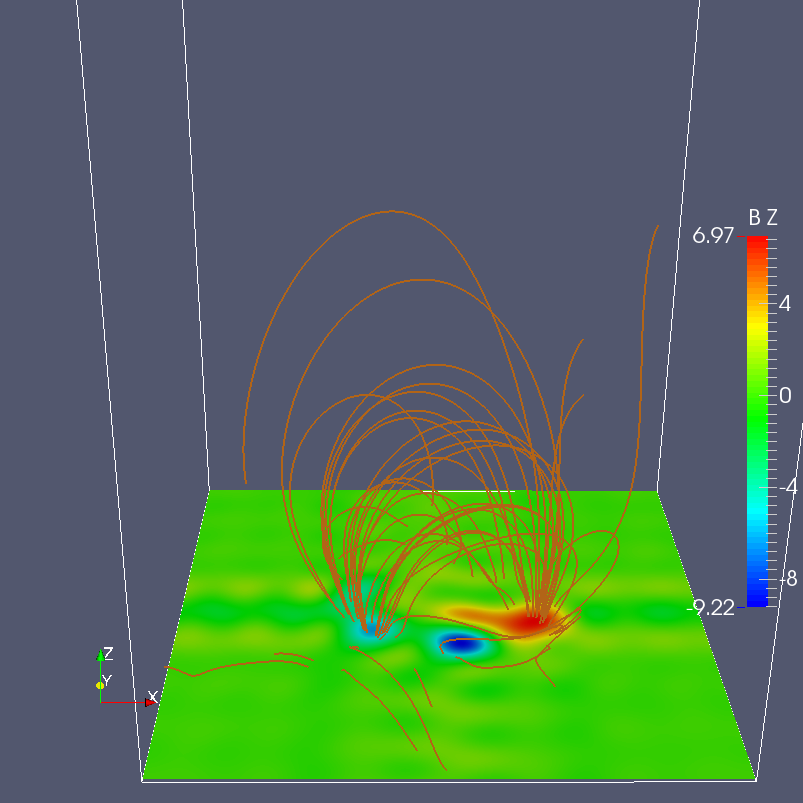}
\caption{Initial structure of the magnetic field in the parallel scaling
tests. The field is current-free and it is extrapolated from the
2D magnetogram of AR1429 by Fourier method. The evolution is triggered by
divergence free velocity vortex located in the magnetic positive
foot-point.}
\label{fig:3Dmagnetic}
\end{figure}

The initial density distribution is prescribed such that the chromospheric
density is 500 times larger than the density in the corona according to the
equation
\begin{equation}
\rho(z)=\frac{\rho_{ch}}{2}\left[1-\tan\left(2\left(z-z_0\right)\right)\right]
+\rho_{co}
\end{equation}
where, $\rho_{ch}$ and $\rho_{co}$ are chromospheric and coronal densities,
respectively. Note that the normalizing density is
$\rho_0 = 2\times10^{15}\,\mathrm{m^{-3}}$.
The transition region is initially localized around $z_0=3$, which corresponds
to 15 Mm.
The initial thermal pressure $p=0.01 p_0 = 0.7957\,\mathrm{Pa}$ is
homogeneous throughout the whole simulation domain, i.e. gravity effects are
neglected. According to the ideal gas law $T=p/(k_BN)$ this reveals the
temperature height profile. The initial density and temperature height
profiles are depicted in Figure~\ref{fig:TemDens}. As one can see in the Figure,
the initial coronal temperature is of the order of $10^6 \ K$.
The initial plasma velocity is zero everywhere in the corona but finite
in the chromosphere.

For the sides of the simulation box, the boundary conditions are set according
to the MHD-equation compatible line symmetry conditions derived by
\citep{Otto+:2007}.
The top boundaries are open, i.e. $\frac{\partial}{\partial z}=0$,
except the normal to the boundary component of the magnetic field
which is obtained to fulfill the source-freeness condition
$\nabla \cdot \boldsymbol{B}=0$.
The bottom boundary of the simulation box is open for entropy and
magnetic fluxes.

The coronal plasma is driven via a coupling to the neutral gas below the
transition region. The neutral gas is driven in accordance with the
observed photospheric motion. First, the plasma flow velocities are inferred
from photospheric magnetic field observations according
to~\citep{Santos:2008-2}.
In order to avoid emerging and submerging magnetic fluxes the motion pattern
is then modeled by divergence-free vortices given by

\begin{equation}
 \boldsymbol{u}_0=\nabla\times\left[ 
\frac{\phi_0}{\cosh\left(\frac{x-y+c_0}{l_0}\right)\cosh\left(\frac{x+y+d_0}{l_1
}\right)} \right]\hat{\boldsymbol{z}}
\end{equation}

The parameters determining strength and localization of the vortex
motion are chosen in accordance with observations.
In the simulated case the magnetic fluxes rotate around footpoints given
by the set of parameters $\phi_0=0.1$, $c_0=9$, $d_0=-49$, $l_0=2$, and
$l_1=-2$.
The strength of the plasma driving by the neutral gas is decreasing with
the height above the photosphere. This decrease is controlled by a
height-dependent coupling term in the momentum equation
Eq.~(\ref{eq:momentum}) (or Eq.~\ref{eq:momentum2}).
The height dependent collision coefficient is defined as

\begin{equation}
 \nu(z)=\frac{\nu_0}{2}\left[1-\tanh(20(z-z_c))\right]
\end{equation}

For the simulated case a good approximation for the coupling coefficient
is $\nu_0=3$ with and $z_c=0.25$ (or 1.25 Mm) as the characteristic height,
where the coupling (and, therefore, the photospherically caused plasma
driving) vanishes.

\subsection{Computational performance of GOEMHD3}
\label{scalability}

Employing the physical setup (i.e.\ initial and boundary conditions) described 
in the previous subsection, the parallel scalability and efficiency of the 
GOEMHD3 code was assessed across a wide range of CPU-core counts and for 
different sizes of the numerical mesh.
The benchmarks were performed on the high-performance-computing system of the 
Max Planck Society, "Hydra", which is operated by its computing centre, RZG. 
Hydra is an IBM iDataPlex cluster based on Intel Xeon E5-2680v2 "Ivy Bridge" 
processors (2 CPU sockets per node, 10 cores per CPU socket, operated at 
$2.8$~GHz) and an InfiniBand FDR~14 network. Hydra's largest partition with a 
fully nonblocking interconnect comprises 36\,000 cores (1800 nodes). For the 
benchmarks Intel's FORTRAN compiler (version 13.1) and runtime were used 
together with the IBM parallel environment (version 1.3) on top of the Linux 
(SLES11) operating system.

\begin{figure}[!ht]
\centering
\includegraphics[width=0.9\hsize]{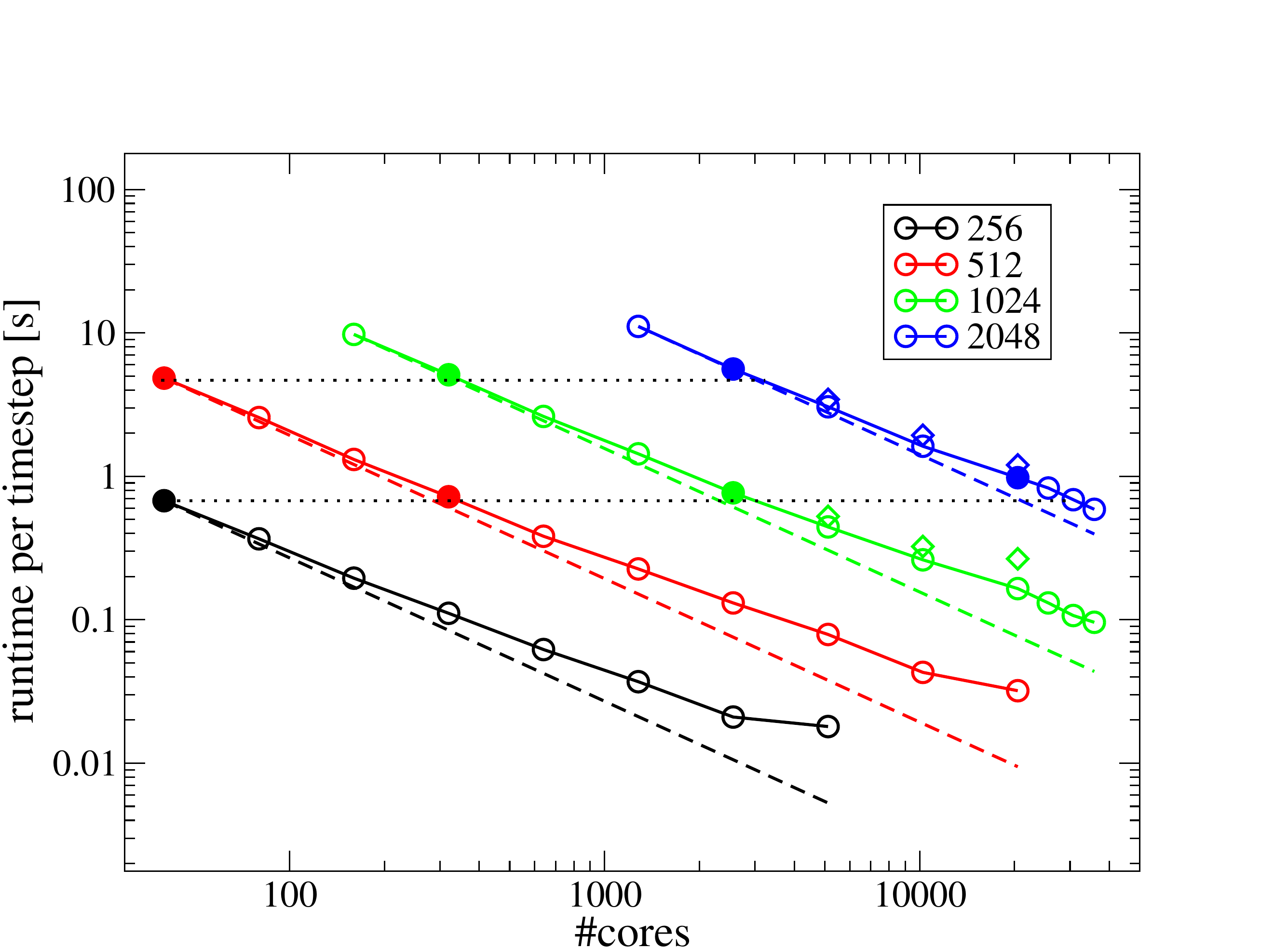}
\caption{Computing time per timestep (open circles) as a function of the number 
of CPU cores. Two MPI tasks, each spawning $10$ OpenMP threads which are
assigned to the 10 cores of a CPU socket were used on each of the 2-socket 
nodes, i.e.\ the total number of MPI tasks is ten times smaller than the number 
of
CPU cores given on the abscissa. Different colours correspond to different 
sizes 
of the numerical grid. Dashed inclined lines indicate ideal \emph{strong}
scalability for a given grid size. Two sets of measurements in which both, the 
number of grid points and the number of processor cores, was increased by a
factor of $2^3$ from left to right are marked by filled circles. The horizontal 
dotted lines are a reference for ideal \emph{weak} scalability. The diamond
symbols correspond to additional runs which employed a plain MPI 
parallelization 
(OpenMP switched off), i.e.\ the number of MPI tasks equals the number of
cores.}
\label{fig:fig08}
\end{figure}

Figure~\ref{fig:fig08} provides an overview of the parallel performance of 
GOEMHD3, using the execution time for a single timestep \footnote{Although in 
principle the runtime per timestep can vary in the course of a simulation due 
to 
the smoothing algorithms being activated in different regions of the grid the 
actual variations are negligible in practice.} as a metric. Four different grid 
sizes are considered, namely grids with $256^3$ cells (black colour), $512^3$ 
cells (red), $1024^3$ cells (green) and $2048^3$ cells (blue). The figure 
demonstrates a very good overall \emph{strong scalability} of the code, i.e. 
the 
reduction of the computing time for fixed grid size with an increasing number 
of 
CPU cores (compare the measured runtimes plotted as circles with the dashed 
lines of the same colour which indicate ideal scalability). For example, the 
parallel efficiency is at the $80\%$ level for the $1024^3$ grid on 2580 cores 
(128 nodes) when compared to the baseline performance on 160 cores (8 nodes). 
Simulations with a $2048^3$ grid can be performed with a parallel efficiency of 
$80\%$ on more than $10\,000$ cores.

Increasing the number of grid points by a factor of 8 (from $256^3$ to $512^3$, 
or from $512^3$ to $1024^3$) and at the same time using an eightfold number of 
CPU cores, the computing time remains almost constant (compare the two sets of 
filled circles with the corresponding horizontal dotted lines in 
Figure~\ref{fig:fig08}). This demonstrates a very good \emph{weak scalability} 
of GOEMHD3, given that the complexity of the algorithm scales linearly with the 
number of grid points.

The deviations from the ideal (strong) scaling curves which become apparent at 
high core counts are due to the relatively larger fraction of time spent in the 
MPI communication (halo exchange) between the domains. For example, for the 
$1024^3$ grid, the percentage of communication amounts to $30\%$ for $10\,000$ 
cores and increases up to about $50\%$ at $36\,000$ cores.
For a given number of cores, the communication-time share is larger for smaller 
grids (manifest as a larger deviation from ideal scalability in 
Figure~\ref{fig:fig08}). The latter observation underlines the benefit of 
making 
the MPI domains as large as possible which is enabled by our hybrid MPI-OpenMP 
parallelization approach (cf.\ Sect.~\ref{Parallelization}). Moreover, by 
comparison with runs where the OpenMP parallelization was switched off and 
compute nodes were densely populated with MPI tasks (one  MPI task per core), 
the advantages of the hybrid MPI-OpenMP  vs.\ a "plain" MPI parallelization 
become immediately apparent. The smaller size of the MPI domains in the "plain" 
MPI runs (diamond symbols in Figure~\ref{fig:fig08}) accounts for a larger 
communication-to-computation ratio and a larger number of smaller MPI messages. 
Accordingly, the communication times increase by about $75\%$, resulting in 
total runtimes being larger by $15-30\%$ when compared to the hybrid version 
using the same number of cores.
It has to be noted, that it is crucial for the hybrid approach to achieve a 
close-to-perfect parallel efficiency of the OpenMP parallelization within the 
MPI domains in order not to jeopardize the aforementioned performance 
advantages 
of the more efficient communication. Additional benchmarks have shown that 
GOEMHD3 indeed achieves OpenMP efficiencies close to 100\% up to the maximum 
number of cores a single CPU socket provides (10 cores on our benchmark 
platform), but -- due to the effects of NUMA\footnote{non-uniform memory 
access} 
and limited memory bandwidth -- not beyond.

Overall, GOEMHD3 achieves a floating-point performance of about $1$~GFlops/s 
per 
core which is about $5\%$ of the theoretical peak performance of the Intel Xeon 
E5-2680v2 CPU. Floating-point efficiencies in this range are commonly 
considered 
reasonable for this class of finite-difference schemes.

\subsection{3D simulation of the energy distribution in the photospheric driven 
solar corona}
\label{results}

In order to understand the dependence of the energy distribution in the
corona on the inflow of mechanical, thermal and magnetic (Poynting flux)
energy from below, through the transition region, we calculated the
corresponding coronal energy contents and the fluxes through the
transition region.

The energies are calculated based on their rates of change as

\begin{eqnarray}
 E_{kin} = \int \left[ -\frac{1}{2}\int_{S}\rho u^2 \boldsymbol u \cdot \mathrm 
d\boldsymbol S \right. \nonumber \\  \left.
  -\frac{1}{2} \int_{V} \left( \boldsymbol u\cdot \boldsymbol \nabla p + 
\boldsymbol u \cdot \boldsymbol j \times \boldsymbol B\right) \mathrm d V 
\right]
\mathrm d t\\
 E_{mag}= \int \left[ \int_{S}\left(-\boldsymbol u B^2 + \left( \boldsymbol u 
\cdot \boldsymbol B\right)\boldsymbol B- \eta \boldsymbol j \times \boldsymbol B
\right) \cdot \mathrm d \boldsymbol S \right. \nonumber \\  \left.
+\int_V \left( -\boldsymbol u \cdot \boldsymbol j \times \boldsymbol B - \eta 
j^2 \right)\mathrm d V \right] \mathrm d t \\
 E_{th}= \int \left[ \frac{-\gamma}{2(\gamma-1)}\int_S p \boldsymbol u \cdot 
\mathrm d \boldsymbol S \right. \nonumber \\  \left.
 + \frac{1}{2} \int_V \left(\boldsymbol u \cdot \nabla p + \eta j^2 
\right)\mathrm d V \right] \mathrm d t
\end{eqnarray}

Note that the main contributions to the surface integrals ($\int_{S}$) are 
mainly due to energy fluxes through the transition region. The latter is taken 
as the lower boundary for the volume integrals ($\int_{V}$). At the same time 
the energy fluxes through the side boundaries cancel each other due to the 
symmetric boundary conditions and the fluxes through the upper boundary are 
negligibly small.

\begin{figure}[!htp]
\centering
\includegraphics[width=0.7\hsize,angle=-90]{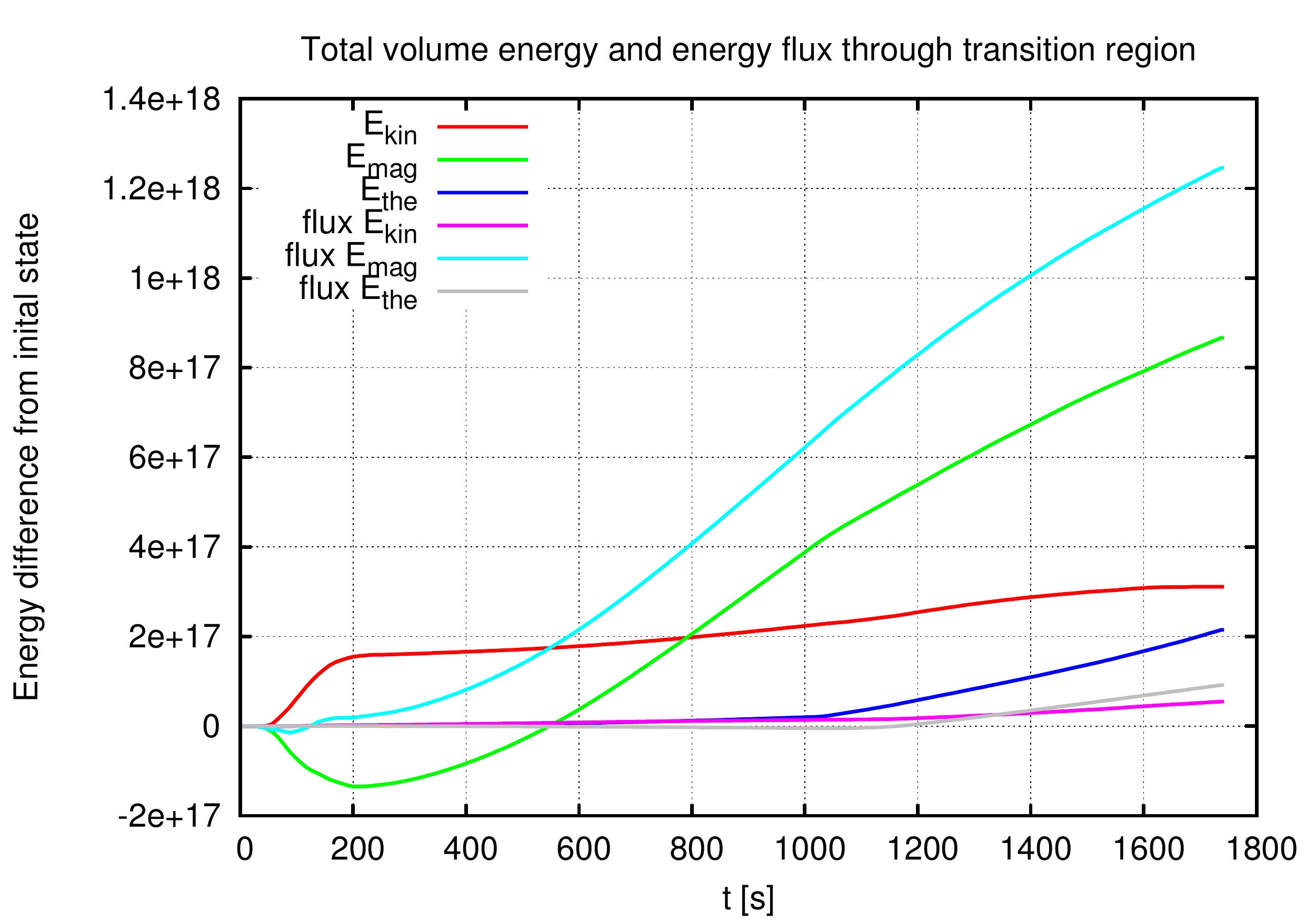}
\caption{Scaling test simulating the solar corona above AR 1429. Shown are the 
temporal evolution of thermal, kinetic and magnetic energies within corona 
above 
the transit region. The energy fluxes of the thermal, kinetic and magnetic 
energies from the chromosphere are denoted by \textit{flux}. For the meaning of 
the different lines see the line form legend. After $t\sim 16$ minutes the 
background resistivity is enhanced causing Joule heating.}
  \label{fig:difEngTF}
\end{figure}

In order to investigate the dependence of the energy distribution on the 
dissipative properties of the coronal plasma we start imposing the 
photospheric-chromospheric driving on an as usual large-Reynolds-number (weakly 
dissipative) corona. Hence the simulation is initiated with a very small 
background resistivity $\eta=10^{-10}$.
According to our normalization length the corresponding characteristic Reynolds 
number based on the normalizing Alfv\'en speed, i.e. the Lundquist number, is 
of 
the order of $10^{10}$; at the grid resolution scale it is still $2 \times 
10^9$. After $t=100\tau_A$ ($\sim1025$~s), when enhanced activity was observed 
at the Sun, the background  resistivity is enhanced to $\eta=10^{-2}$ which 
corresponds to microturbulence theory predictions \cite{SilinBuchner2003-3}, 
\cite{SilinBuchner:2003-4}. Figure~\ref{fig:difEngTF} depicts the temporal 
evolution of the kinetic, magnetic and thermal energies within the corona above 
the transition region and the energy fluxes into/from the corona across the 
transition region. Note that the curves in the figure correspond to the net 
changes of the energy, i.e.,  the excess from the initial values at $t=0$. The 
figure shows that main energy source for the corona is the Poynting flux 
generated by the footpoint motion of the flux tubes, not the direct transfer of 
kinetic energy from the chromosphere. Until about $t = 20\tau_A$ (about $200 \ 
s 
$, a little more than $3 \ min $) the magnetic energy inflow is enhancing 
mainly 
the kinetic energy of the corona, i.e. the coronal flux tubes are driven by the 
photospheric motion. This process lasts as long as the average propagation time 
of the corresponding Alfvenic perturbation along typical flux tubes. Hence this 
Alfven transition time is needed to drive, finally, the whole flux tube system. 
After that the amount of magnetic energy in the corona steadily increases 
until, 
at $t=100\tau_A$, i.e. after about 1000~s (i.e. about 17 min) the resistivity 
is 
increased by orders of magnitude (see above). Now the enhanced resistivity (= 
magnetic diffusivity) quickly heats the corona. Already after only $80\tau_A$ 
(800 s or 13 min) the thermal energy enhancement of the corona due to the 
imposed Joule heating reaches almost the level of the kinetic energy 
enhancement 
due to the footpoint motion. At the same time the increase of the magnetic 
energy contents of the corona due to the permanent Poynting flux inflow is 
slowed down only slightly by the heating process.

\begin{figure}[!htp]
\centering
\includegraphics[width=0.7\hsize,angle=-90]{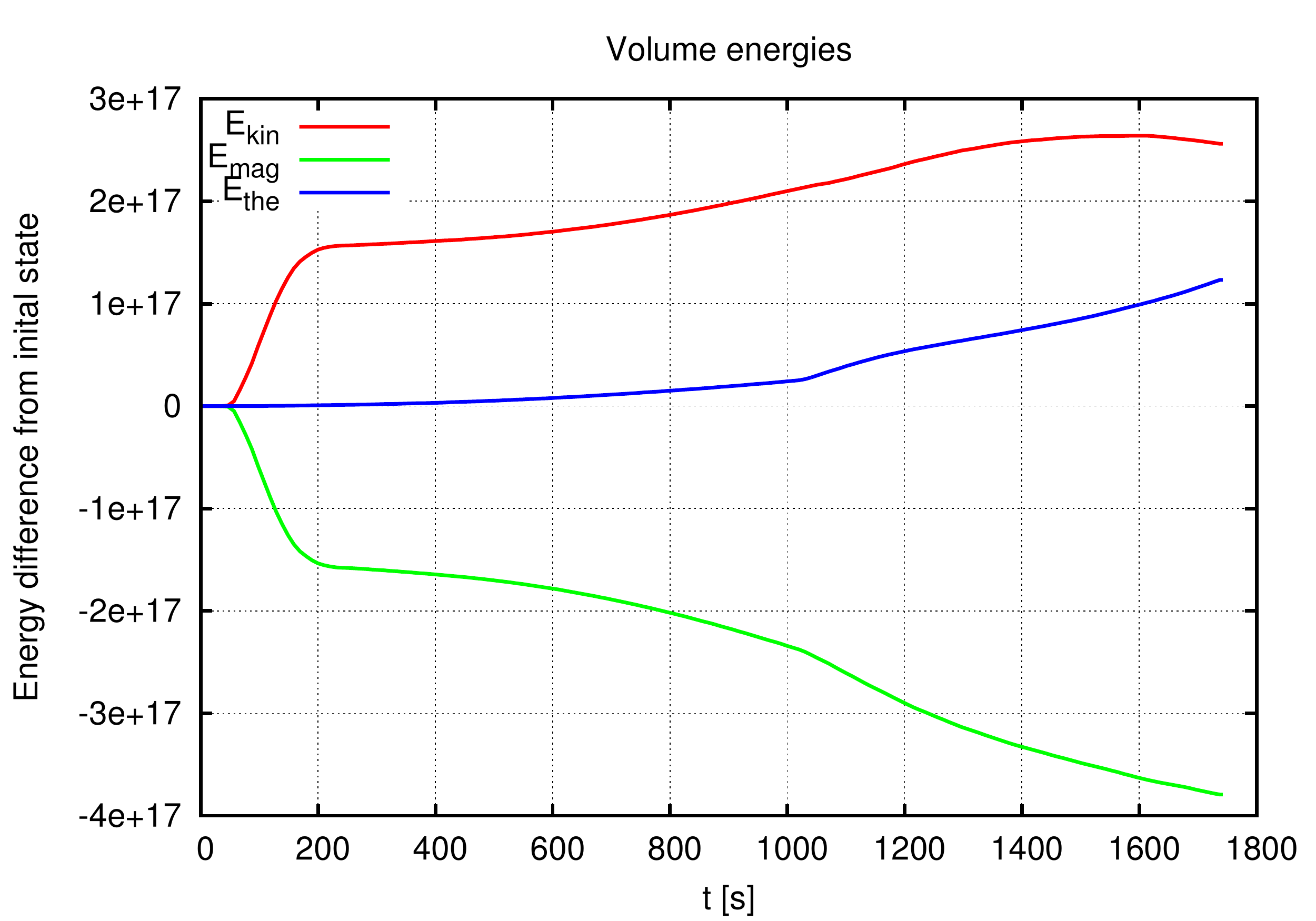}
\caption{Scaling test simulation of the solar corona above AR 1429: temporal 
evolution of the thermal, kinetic and magnetic volumetric energies in the solar 
atmosphere above the transition region. For the meaning of the different lines 
see the line form legend. Note that after the time $t \sim 16 minutes$~s the 
resistivity and, therefore, Joule heating is essentially enhanced.}
  \label{fig:difEngVol}
\end{figure}

For a better understanding of the change of the coronal energy distribution
Figure~\ref{fig:difEngVol} depicts the temporal evolution of its kinetic,
magnetic and thermal energy contents without taking into account the
contribution of energy inflows across the transition region.
As one can see in the Figure first, after the Alfvenic transition time
has passed, in the course of the almost ideal (large Reynolds number)
evolution practically only the kinetic energy of the corona grows completely
at the expense of the decreasing magnetic field energy. Then, after the
magnetic diffusivity is enhanced at  $t=100\tau_A$, the magnetic energy
drops faster due to resistive dissipation. The latter enhances the thermal
energy contents of the corona via Joule heating. After $t=170\tau_A$ the
amount of the released energy within the corona is about half of the kinetic
energy as one already could see in Figure~\ref{fig:difEngTF}.

The time evolution of the magnetic field is captured by a movie
which can be obtained on the WWW
\footnote{http://www.mpg.de-streaming-eu.s3.amazonaws.com/ 
de/institute/mps/magnetic\_field\_AR11429\_buechner.mp4}.

\begin{figure*}[!htp]
\centering
\includegraphics[width=0.6\hsize,bb=0 0 944 777]{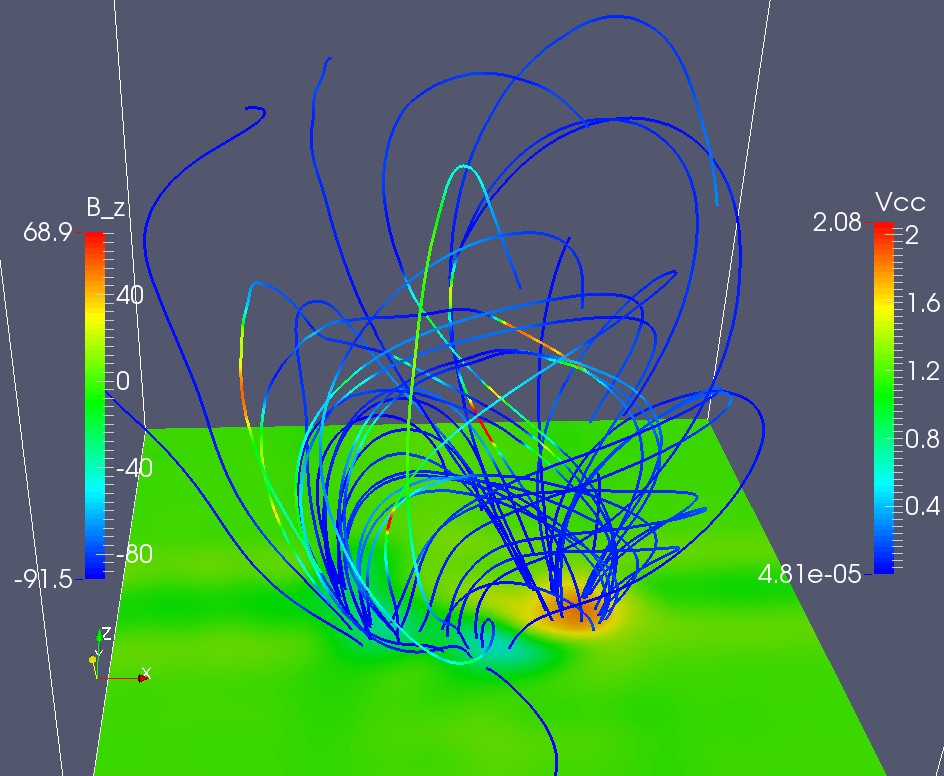}
\caption{Snapshot of the magnetic field at the time $t=130\tau_A$
($\sim 22$ Minutes) of the simulated AR 1429.
The magnetic field lines are coloured by the magnitude of the current carrier
velocity $j/n$.
The bottom plane depicts magnetic field $B_z$ component
(perpendicular to the plane).}
\label{fig:Bevolt130}
\end{figure*}

The movie shows that until the moment when the resistivity, i.e. the magnetic 
diffusivity is enhanced (after about 16 minutes solar time) the coronal 
magnetic 
fields evolves almost ideally, just being bended following the footpoint motion 
while the magnetic flux tubes are kept low. Only after the (colour coded along 
the field lines) current carrier velocity $ j / n$ becomes, large enough, i.e. 
after the micro-turbulence threshold is reached, the flux tubes start to rise 
faster. The reason is that the enhanced current dissipation allows magnetic 
diffusion and heating. After that the magnetic flux tubes continue to rise even 
faster releasing parts of the high magnetic tension until, finally, 
reconnection 
starts, the most efficient magnetic energy release process. 
Figure~\ref{fig:Bevolt130} shows the magnetic field configuration reached at 
$t=130\tau_A$, i.e. after about $22$~minutes. The colour coding of the magnetic 
field lines depicts the actual values of $V_{CC} = j / n$. At the places where 
$V_{CC}$ is enhanced above the threshold the plasma is quickly heated by Joule 
current dissipation.

\subsection{Estimate of the numerical dissipation}
\label{sec:numDiss}
\changed{
Using a simulation setup for an eruptive solar region similar to the one 
described in Sect.~\ref{conditions}, we give an estimate of the effective 
numerical dissipation rate for the Leap-Frog scheme in realistic, 
three-dimensional simulations with complete input physics (see also 
Sect.~\ref{Harris_sheet}). The mesh resolution is set to $258^3$ grid points. 
The initial magnetic field was taken from the active region 11226 on $7^{th}$ 
June 2011 06:16 UT. During a total simulation time of $t=2400$ and about 
$9.66\times10^5$ time steps the field line apex rose from an initial altitude 
of 
$z_0=3.2$ to the final height, $z_e=29.0$. Due to the very low physical 
resistivity the field line is frozen in to the plasma and one can predict the 
position of the second foot-point by tracking photospheric plasma motions. In 
our simulation the displacement of the second foot-point from the predicted 
position was  $\Delta r_e=0.02942$ from which we can estimate effective 
numerical resistivity at the end of the simulation:
\begin{equation}
 \eta_n=\frac{\Delta r_e^2}{\Delta 
t}=\frac{0.02942^2}{2400}\approx3.6\times10^{-7}
\end{equation}
The corresponding value of $\eta_n$ for the times $t=20$ and $t=240$ are 
$1.47\times10^{-8}$ and $1.14\times10^{-7}$, respectively. Different values for 
$\eta_n$ are obtained for different times because the computation of the 
foot-point displacement includes errors from the field line integration (which 
is very long at the end of the simulation) and also the from tracing the second 
foot-point over time. In summary, both the idealized, one-dimensional Harris 
current sheet (Sect.~\ref{Harris_sheet}) and also the application of GOEMHD3 to 
the full solar corona physics in three dimensions reveal no significant 
reconnection due to numerical dissipation.  

}
\section{Discussion and conclusions}
\label{Conclusions}

We have implemented a new, three-dimensional MHD code based on 
second-order-accurate finite-difference discretization schemes in order to be 
able to efficiently simulate large-scale weakly-dissipative 
(large-Reynolds-number) astrophysical plasma systems. In order to 
\changed{reduce} numerical dissipation the conservative part and source terms 
of 
the equations are solved by a Leap-Frog scheme which is second order accurate 
in 
time and in space. Only terms with second order spatial derivatives, i.e. 
viscosity, diffusion and resistive dissipation, are discretized by a 
DuFort-Frankel scheme. Numerically induced grid-scale oscillations are damped 
away by introducing an artificial diffusivity which is switched on locally. In 
this paper we have documented the main physical, numerical and computational 
concepts of the new GOEMHD3 code as well as its computational performance. The 
code was tested and verified by means of a number of appropriate test problems 
which allowed us to reveal the limits of the applicability of GOEMHD3 and to 
describe the ways to achieve the goals when solving concrete problems.

First the code was tested by simulating a velocity-shear Kelvin-Helmholtz 
instability. Owing to the use of \changed{a low numerical dissipation} 
Leap-Frog scheme GOEMHD3 obtained the same linear evolution as simulations by 
the numerically more expensive, higher order PENCIL code. As expected, at later 
times, during the non-linear evolution of the instability for the same number 
of 
grid points, the dissipation is larger than that of the higher-order PENCIL 
code 
\citep{McNally+:2012}. The reason is artificial diffusivity which is locally 
switched on in order to damp spurious grid-scale oscillations inherent to the 
Leap-Frog scheme. The amount of necessary damping can, however, easily and 
computationally cheaply be reduced by enhancing the grid resolution of the 
overall less expensive second-order scheme. We showed that GOEMHD3 solutions 
converge towards PENCIL's solution and the result uncertainty (GCI, 
Figure~\ref{fig:GCI}) is in good agreement with the relative error 
(Figure~\ref{fig:errMod}) of the GOEMHD3's (compared to PENCIL) mode amplitude 
evolution. GOEMHD3 revealed the same results for \citet{Orszag+Tang:1979} 
vortices as obtained by \citet{Ryu+:1995} and by \citet{Dai+Woodward:1998}. 
Gradients are well resolved by two grid-points. Numerical oscillations are 
smoothed away by locally switching on diffusivity. GOEMHD3 dissipates more 
energy at steep wave fronts as compared to a higher-order code for the same 
grid-resolution. This dissipation can be easily overcome by (locally) using a 
larger number of grid points.
The solver for the resistive part of the induction equation was tested 
separately by imposing a homogeneous resistivity on a current column. The 
results are in good agreement with an analytically predicted current decay. The 
code fully reproduces the analytic solution until the enhanced numerical errors 
reach the center of the current system where the current concentration is 
maximum. Since the spreading of the numerical error depends only on the number 
of time steps, not on the real physical time, this phenomenon is of purely 
numerical nature. In order to cope with this effect GOEMHD3 contains a module 
which smooths an eventually self-regulated resistivity increase around the 
maximum gradient of the current growth. We showed that such resistivity 
smoothing is sufficient to keep simulations stable.
\changed{
In simulations of a one-dimensional Harris current sheet and of a realistic, 
three-dimensional scenario with complete input physics the residual numerical 
dissipation of GOEMHD3 was demonstrated to be sufficiently small to allow 
applications with almost ideal magneto-fluids at very high Reynolds number 
($Re\sim 10^{10}$). 
}

The parallel computing performance of the code was demonstrated by obtaining 
the 
scaling of the runtime with the number of CPU cores and grid points (i.e.\ 
different numerical resolutions) for a realistic application scenario. To this 
end GOEMHD3 was initialized to simulate the evolution of the solar atmosphere 
above an observed active region and thus to obtain the distribution of the 
energy injected from the photosphere through the transition region into the 
corona. The calculations revealed an almost linear \emph{ strong scaling} of 
the 
runtime with the number of CPU cores for meshes with up to $2048^3$ grid 
points. 
On the HPC system \emph{Hydra} of the Max-Planck Society GOEMHD3 exhibited an 
almost ideal scaling even beyond 30\,000 processor cores. In addition, also a 
very good {\it weak scalability} from 20 cores (1 node) for $256^3$ grid runs 
to 
more than 20\,000 cores (1000 nodes) for a $2048^3$ grid, was obtained, thereby 
maintaining absolute run times of less than a second per time step.

In summary, we have shown that the new GOEMHD3 code is able to efficiently and 
accurately solve the MHD equations of almost ideal plasma systems on 
non-equidistant grids. Due to its second-order-accurate discretization scheme 
the code is conceptually straightforward to implement and to parallelize on 
distributed-memory computer architectures. The code can simply be adjusted to 
different types of initial and boundary conditions and extended to include 
additional physics modules. Due to its excellent computational performance and 
parallel efficiency the formally comparably low numerical accuracy per grid 
point and time step can easily be compensated by adopting an enhanced 
resolution 
in space and time. Aiming at the same accuracy for the same problem this is 
computationally still cheaper than running codes using higher-order schemes.

\begin{acknowledgements}
We gratefully acknowledge the support of this work by the German Science
Foundation \emph{Deut\-sche For\-schungs\-ge\-mein\-schaft, DFG\/}, project CRC
963-A2, by the Czech GACR project 13-24782S and the EU FP7-PEOPLE-2011-CIG
programme PCIG-GA-2011-304265 (SERaF) as well as by Dr. Bernhard Bandow at the
Max-Planck Institute for Solar System Research. The authors thank Colin P.
McNally who provide us data for the cross-code comparison and an anonymous
referee for valuable comments that helped to improve the quality of the paper.
\end{acknowledgements}

\bibliographystyle{aa}
\bibliography{20141105}

\end{document}